\documentclass[twocolumn,preprintnumbers,amsmath,amssymb,superscriptaddress,pre,showpacs]{revtex4}
\usepackage[dvips]{graphicx}
\usepackage{dcolumn}
\usepackage{bm}
\usepackage[dvips]{epsfig}
\newcommand{\gp}{\dot{\gamma}}
 
\newcommand{\gloc}{\dot{\gamma}_{\rm loc}}
\newcommand{\geff}{\dot{\gamma}_{\rm eff}}
\newcommand{\gps}{\dot{\gamma}^\star}

\begin{document}

\title{Yielding dynamics of a Herschel-Bulkley fluid: a critical-like fluidization behaviour}

\author{Thibaut Divoux}
\affiliation{Universit\'e de Lyon, Laboratoire de Physique, \'Ecole Normale Sup\'erieure de
Lyon, CNRS UMR 5672, 46 All\'ee d'Italie, 69364 Lyon cedex 07, France.}
\author{David Tamarii}
\affiliation{Universit\'e de Lyon, Laboratoire de Physique, \'Ecole Normale Sup\'erieure de
Lyon, CNRS UMR 5672, 46 All\'ee d'Italie, 69364 Lyon cedex 07, France.}
\author{Catherine Barentin}
\affiliation{Laboratoire de Physique de la Mati\`ere Condens\'ee et Nanostructures, Universit\'e de Lyon; Universit\'e Claude Bernard Lyon I, CNRS UMR 5586 - 43 Boulevard du 11 Novembre 1918, 69622 Villeurbanne cedex, France.}
\author{Stephen Teitel}
\affiliation{Department of Physics and Astronomy, University of Rochester, Rochester, NY 14627, USA.}
\author{S\'ebastien Manneville}
\affiliation{Universit\'e de Lyon, Laboratoire de Physique, \'Ecole Normale Sup\'erieure de
Lyon, CNRS UMR 5672, 46 All\'ee d'Italie, 69364 Lyon cedex 07, France.}
\affiliation{Institut Universitaire de France.}
\date{\today}

\begin{abstract}
The shear-induced fluidization of a carbopol microgel is investigated during long start-up experiments using combined rheology and velocimetry in Couette cells of varying gap widths and boundary conditions. As already described in [Divoux et al., {\it Phys. Rev. Lett.}, 2010, {\bf 104}, 208301], we show that the fluidization process of this simple yield stress fluid involves a transient shear-banding regime whose duration $\tau_f$ decreases as a power law of the applied shear rate $\gp$. Here we go one step further by an exhaustive investigation of the influence of the shearing geometry through the gap width $e$ and the boundary conditions. While slip conditions at the walls seem to have a negligible influence on the fluidization time $\tau_f$, different fluidization processes are observed depending on $\gp$ and $e$: the shear band remains almost stationary for several hours at low shear rates or small gap widths before strong fluctuations lead to a homogeneous flow, whereas at larger values of $\gp$ or $e$, the transient shear band is seen to invade the whole gap in a much smoother way. Still, the power-law behaviour appears as very robust and hints to critical-like dynamics. To further discuss these results, we propose (i) a qualitative scenario to explain the induction-like period that precedes full fluidization and (ii) an analogy with critical phenomena that naturally leads to the observed power laws if one assumes that the yield point is the critical point of an underlying out-of-equilibrium phase transition.
\end{abstract}
\maketitle

\section{Introduction}

A huge amount of soft materials display solid properties at rest and under relatively small shear stresses while they become fluid above a typical stress known as the ``yield stress.''\cite{Barnes:1999} Examples range from daily products such as the foams or emulsions encountered in foods and cosmetics to drilling muds and granular materials. \cite{Hohler:2005,Weeks:2007,Schall:2010} This property is also shared by both hard and soft jammed colloids which have been used as model yield stress fluids.\cite{Coussot:2007} Besides intense debate over the right experimental procedure to measure the yield stress\cite{Nguyen:1992,Moller:2006} --or even on its very existence\cite{Roberts:2001,Moller:2009a}--, the question of {\it how} a yield stress material turns from solid to liquid under shear remains largely unexplored. In fact, previous works have mainly focused on the steady state reached by the material after yielding, with emphasis on whether or not the stationary flow field displays shear banding, i.e. whether an unsheared solid region coexists with a fluidized region characterized by a finite local shear rate at steady state or whether the whole sample flows homogeneously with a finite shear rate.\cite{Derec:2001,Coussot:2002a,Picard:2002,Dennin:2008,Moller:2008,Coussot:2010,Mansard:2011} This has recently led to distinguish between ``simple'' yield stress fluids which do not show steady-state shear banding, and the other yield stress fluids which display solid--fluid coexistence at steady-state.\cite{Ragouilliaux:2007,Moller:2009b} On the one hand, simple yield stress fluids are made of soft repulsive particles and encompass foams\cite{Ovarlez:2010}, emulsions, and some microgels such as the carbopol dispersions considered in the present work. Their steady-state rheology is well described by a Herschel-Bulkley model\cite{Roberts:2001,Salmon:2003a,Hohler:2005}: $\sigma=\sigma_c+\tilde\eta\gp^n$, where $\sigma$ is the shear stress, $\sigma_c$ is the yield stress, $\gp$ is the shear rate, and $\tilde\eta$ and $n=0.2$--$1$ are phenomenological parameters. On the other hand, the rest of the yield stress materials, which includes for instance soft colloidal glasses, clay suspensions and attractive emulsions,\cite{Becu:2006,Ragouilliaux:2007,Rogers:2008} display steady-state shear banding. Such permanent flow heterogeneities are commonly interpreted as the result of mechanical instabilities that derive from underlying non-monotonic flow curves,\cite{Sollich:1997,Sollich:1998,Varnik:2003,Berthier:2003,Coussot:2010,Mansard:2011} of thixotropic behavior,\cite{Ragouilliaux:2007,Ovarlez:2009,Mewis:2009} or more recently of a coupling between shear and concentration \cite{Besseling:2010}. In this latter case, tiny local concentration variations can result locally in large variations of viscosity and thus in flow heterogeneities. 
However, such a simple distinction between simple and other yield stress materials may turn out to be experimentally quite difficult to test, (i) because of vanishingly small flow velocities and shear rates close to yielding, (ii) because of transient regimes that can become critically long, and (iii) because the boundary conditions\cite{Gibaud:2008,Gibaud:2009} and/or stress initial conditions\cite{Cheddadi:2011} may affect the steady state behaviour. To overcome these difficulties, spatially and temporally resolved measurements performed for various boundary conditions are required on time scales long enough to ensure that a steady state has been reached.

In order to assess the slow yielding dynamics of a simple yield stress fluid, we have recently used ultrasonic velocimetry coupled to standard rheology in carbopol samples sheared in a small-gap concentric cylinder cell (Couette cell).\cite{Divoux:2010} We have shown that, under a given imposed shear rate, the transition from solidlike to fluidlike behaviour involves a transient regime characterized by shear banding: a fluidized band is generated at the inner rotating cylinder and, after a time $\tau_f$ called the ``fluidization time'' that can reach $10^5$~s, it invades the whole gap of the Couette cell. $\tau_f$ was shown to follow a power-law dependence on the imposed shear rate: $\tau_f\sim \gp^{-\alpha}$ with $\alpha=2$--3, independent of the shearing geometry but depending on the protocol for sample preparation and on the carbopol concentration. At short times, the shear band grows from a thin lubrication layer that is generated at the inner wall as the sample suddenly fails after a period of elastic loading, for both rough and smooth boundary conditions. This short-time behaviour, which has been described in detail elsewhere\cite{Divoux:2011b}, is fully correlated with the presence of a stress overshoot in the rheological response of the material.

Finally, experiments under controlled shear stress\cite{Divoux:2011a} have shown that a similar fluidization scenario is at play during creep tests above the yield stress and that the fluidization time follows a power law of the viscous stress $\sigma-\sigma_c$: $\tau_f\sim (\sigma-\sigma_c)^{-\beta}$ with $\beta=4$--6. By comparing the power laws obtained under imposed shear rate and shear stress, one recovers the Herschel-Bulkley behaviour, in which the exponent $n$ naturally appears as the ratio of the two fluidization exponents $n=\alpha/\beta$. This leads to an original point of view linking the steady-state rheology of the material to transient fluidization processes characterized by critical-like scalings. Note that such a point of view is supported by two other recent experimental works on foams \cite{Katgert:2009} and gels of type-I collagen \cite{Gobeaux:2010} which emphasize the potentially broad interest of our results obtained on a particular carbopol microgel.

The present article is meant to complete our previous experimental work under controlled shear rate\cite{Divoux:2010} and to propose various possible interpretations of our observations. The paper is structured as follows. Section~\ref{s.exp} describes the materials and methods used to investigate yielding of carbopol dispersions. Our results are gathered in Sect.~\ref{s.res}. We first check carefully from both global and local rheological data that {\it at steady state}, our carbopol samples display the hallmark of simple yield stress fluids, namely a Herschel-Bulkley flow behaviour together with homogeneous velocity profiles. We then give a qualitative description of the {\it transient} shear banding phenomenon. The applied shear rate $\gp$ is further varied to evidence the power law followed by the fluidization time $\tau_f$. We also focus on the influence of the shearing geometry on the fluidization dynamics by varying the gap width $e$ and the boundary conditions. Finally, Sect.~\ref{discuss} further discusses the experimental results. We first gather our observations in a ``phase diagram'' for transient shear banding in the plane $(\gp,e)$. We also provide the reader with a qualitative interpretation of the induction-like period that precedes full fluidization at low shear rates. We then show that power laws for the fluidization times can be reproduced from scaling arguments in which the yield point appears as a critical point of an underlying out-of-equilibrium phase transition. Such a general analogy calls for a systematic investigation of the transient fluidization scenario in other yield stress materials.

\section{Experimental}
\label{s.exp}

\subsection{Preparation of carbopol samples}

We focus on aqueous dispersions of carbopol ETD~2050 at a concentration of 1~\%~w/w. Carbopol powder contains homo- and copolymers of acrylic acid highly cross-linked with a polyalkenyl polyether.\cite{Roberts:2001,Baudonnet:2004} Once the polymer is dispersed into water, adequate pH conditions lead to polymer swelling. Swollen polymer particles get jammed into an amorphous assembly referred to as a ``microgel.'' The typical size of the soft particles range from a few microns to roughly 20 microns.\cite{Ketz:1988,Kim:2003,Oppong:2006} The exact microstructure depends on the type of carbopol,\cite{Baudonnet:2004} on its concentration,\cite{Roberts:2001} and on the details of the preparation protocol.\cite{Kim:2003,Baudonnet:2004,Lee:2011} Carbopol microgels are known to be non-aging, non-thixotropic simple yield stress fluids \cite{Piau:2007,Coussot:2009,Moller:2009a,Benmouffok:2010} and their steady-state flow curve nicely follows the Herschel-Bulkley law: 
\begin{equation}
\sigma=\sigma_c+\tilde{\eta}\,\gp^n\,,
\label{eq1}
\end{equation}
with $n=0.3$--0.6 depending on the type of carbopol and its concentration.\cite{Roberts:2001,Divoux:2010,Oppong:2006,Coussot:2009}

Our preparation protocol has been described in detail elsewhere.\cite{Divoux:2011a,Divoux:2011b} As already noted, due to variations of the final pH ($6.5<$pH$<7.5$), the properties of our samples may vary from batch to batch. Therefore, if one wants to compare quantitatively results obtained in various geometries, gaps, or boundary conditions, one should ensure that batches with the same final pH are used. This issue will be mentioned whenever relevant to the present experiments.

It is also important to recall that, in order to perform ultrasonic velocimetry measurements, our samples are seeded with micronsized hollow glass spheres at 0.5~\%~w/w (Potters, Sphericel, mean diameter 6~$\mu$m, density 1.1) to provide acoustic contrast to the samples.\cite{Manneville:2004a} We have shown previously that the addition of these glass spheres has little influence on the sample rheological properties:\cite{Divoux:2011b} it only stiffens the material by about 10~\%. The fact that seeding has no impact on the rheological response of our carbopol microgel during yielding will be further demonstrated in the present study. To this aim, we also prepare 1~\%~w/w ``pure'' carbopol microgels, i.e. samples that are free of acoustic contrast agents.

\subsection{Rheological measurements}
\label{s.rheomes}

Rheological measurements are performed using stress-controlled rheometers (Anton Paar MCR 301 and TA Instruments AR1000N). Here, we shall mainly focused on experiments performed in a polished Plexiglas Couette geometry with a height of 28~mm, a rotating inner cylinder of radius 24 mm, and a fixed outer cylinder of radius 25~mm, yielding a gap width $e=1$~mm. The surface roughness of polished Plexiglas is about 15~nm as measured from atomic force microscopy, which will be referred to as ``smooth'' in the following. Other polished Plexiglas rotors of radii 24.55, 23.5 and 22~mm will be used to vary the gap width, yielding respectively $e=0.45$, 1.5, and 3~mm.

Besides these smooth Couette cells, we will briefly discuss results obtained in a rough Couette cell (height 28 mm, rotating inner cylinder radius 23.5 mm, fixed outer cylinder radius 24.6~mm, gap width $e=1.1$~mm) where sand paper was glued on both shearing surfaces to provide a roughness of 60~$\mu$m. Mixed boundary conditions will also be tested in a Couette cell of height 28 mm and gap width 1.6~mm where only the rotating cylinder was covered with sand paper of roughness 60~$\mu$m while the fixed outer cylinder is the same as for smooth Couette cells. Finally, a rough plate-and-plate geometry (radius 21~mm, gap width $e=1$~mm,  roughness 162~$\mu$m) as well as a smooth aluminum cone-and-plate geometry (radius 25~mm, cone angle 2$^\circ$) will be used to check for the rheological signature of transient shear banding in complementary geometries. 

Before starting an experiment, preshear is applied for 1 min at +1000~s$^{-1}$ and for 1 min at -1000~s$^{-1}$ to erase the loading history.\cite{Cloitre:2000,Viasnoff:2002} The viscoelastic moduli are then monitored for 2~min. We found that both the elastic and the viscous moduli no longer vary significantly after 2 min. Finally, the sample is left at rest for 1 min to ensure that a reproducible initial state is reached. The reader is referred to Ref.~\cite{Divoux:2011b} for more details about the rheological protocol and the viscoelastic properties of our samples at rest.

\subsection{Ultrasonic velocimetry}

In the Couette geometry, velocity profiles across the gap can be recorded with a spatial resolution of 40~$\mu$m using ultrasonic speckle velocimetry (USV). Full technical details about USV can be found in Ref.~\cite{Manneville:2004a}. Here, the sample velocity field is measured at about 15~mm from the cell bottom simultaneously to the global rheological response. This allows for a direct correlation between time-resolved velocimetry and rheological data. The temporal resolution depends on the imposed shear rate and varies from about 50~s per velocity profile at the lowest shear rates ($\gp\lesssim 0.5$~s$^{-1}$) to less than 1~s for $\gp\gtrsim 10$~s$^{-1}$.

\section{Results}
\label{s.res}

\subsection{Herschel-Bulkley behaviour and homogeneous steady-state velocity profiles}
\label{s.locvsglob}

Before addressing the issue of transient fluidization, it is imperative to check that our microgel samples behave as expected for simple yield stress fluids, i.e. (i) that their steady-state flow curve follows the Herschel-Bulkley model and (ii) that the steady-state flow behaviour is characterized by homogeneous velocity profiles consistent with the global rheological data. One way to perform this check is to focus on {\it local} rheology by combining standard rheology and velocimetry in order to plot the {\it local flow curve} $\sigma(r)$ vs $\gp(r)$. Indeed, in the Couette geometry, the local stress reads:
\begin{equation}
\sigma(r) = \sigma_1\left(\frac{R_1}{R_1 + r}\right)^2\, ,
\label{sigmar}
\end{equation}  
the stress at the rotor $\sigma_1$ being given by:
\begin{equation}
\sigma_1 = \frac{\Gamma}{2\pi h R_1^2}\, ,
\label{sigma1}
\end{equation}  
where $\Gamma$ is the torque exerted on the rotor and $h$ is the height of the cell.
On the other hand, the local shear rate can be directly extracted from the velocity profiles through:
\begin{equation}
\gp(r)=-(R_1+r)\,\frac{\partial}{\partial r}\left(\frac{v(r)}{R_1+r}\right)\, .
\label{localshearrate}
\end{equation}
The local flow curve can then be compared to the global flow behaviour. Recently, such an analysis has been used to address the link between local and global behaviours in various systems ranging from emulsions \cite{Salmon:2003a,Becu:2006,Goyon:2008,Ovarlez:2008} and granular pastes \cite{Huang:2005} to industrial materials \cite{Ragouilliaux:2006}. In the case of a simple yield stress fluid and in the absence of strong confinement, it is expected that the local and global data collapse \cite{Ovarlez:2008}. In the following, we focus on the data collected in the smooth Couette geometry of gap width $e=3$~mm since it presents the largest stress heterogeneity, i.e. the largest range of local shear rates and stresses for a given applied shear rate. 

\begin{figure}[!t]\tt
\centering
\includegraphics[width=0.9\columnwidth]{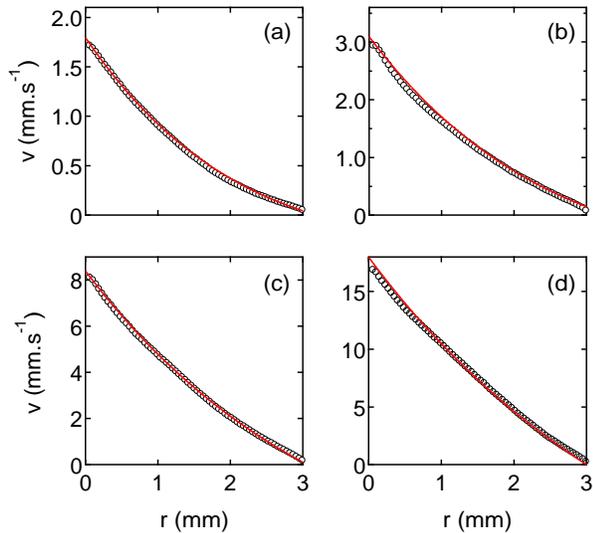}
\caption{Velocity profiles $\langle v(r)\rangle$ averaged over a time window of 150 to 600~s in the steady-state regime for (a) $\gp=0.7$~s$^{-1}$, (b) $\gp=1.2$~s$^{-1}$, (c) $\gp=3.0$~s$^{-1}$, and (d) $\gp=6.0$~s$^{-1}$. The red solid lines are the velocity profiles predicted using the Herschel-Bulkley behaviour derived from the global rheological data of Fig.~\ref{rheol_3mm}. The upper limit of the vertical scale corresponds to the rotor velocity $v_0$. Experiments performed in a smooth Couette cell of gap width 3~mm.}
\label{profiles_3mm}
\end{figure}
\begin{figure}[!t]\tt
\centering
\includegraphics[width=0.9\columnwidth]{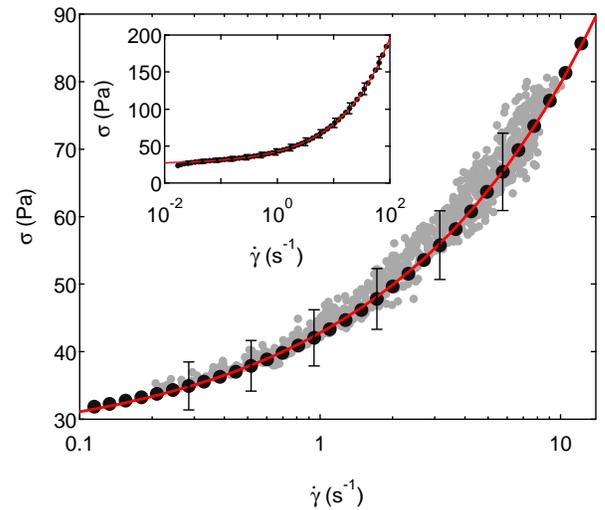}
\caption{Local rheology $\sigma(r)$ vs $\gp(r)$ (gray dots) extracted from the steady-state velocity profiles of Fig.~\ref{profiles_3mm} (see text) and compared to the global rheological data $\sigma$ vs $\gp$ ($\bullet$). The global flow curve was obtained by averaging two independent decreasing shear rate ramps (from $100$~s$^{-1}$ down to $1.7\,10^{-2}$~s$^{-1}$) with a waiting time of 20~s per point and measured respectively before and after the series of start-up experiments under given applied shear rates. The error bars show the difference between the two flow curve measurements. The red line shows the best Herschel-Bulkley fit of the global flow curve for $\gp>3.5\,10^{-2}$~s$^{-1}$: $\sigma=\sigma_c+\tilde\eta\gp^n$, where $\sigma_c=25.7$~Pa, $n=0.50$, and $\tilde\eta=17.1$~Pa\,s$^n$. Inset: full set of global rheological data. Experiments performed in a smooth Couette cell of gap width 3~mm.}
\label{rheol_3mm}
\end{figure}

Figure~\ref{profiles_3mm} gathers a few velocity profiles for $e=3$~mm obtained by averaging over 100 to 500 individual velocity profiles in the steady-state, i.e. when all the fluidization process that will be described later is completed. Depending on the shear rate, this average corresponds to a time span of 150 to 600~s and the steady-state is reached after 10 to $10^3$~s. In all cases, the steady-state velocity profiles are homogeneous and do not present any shear banding. The curvature of the velocity profiles for the lowest shear rates is both due to the rather large stress variation of about 25~\% from the rotor to the stator and to the proximity of the yield stress. As the shear rate is increased and the shear stress departs from the yield stress, velocity profiles become closer to linear.

The local shear rate $\gp(r)$ is easily extracted from steady-state velocity profiles using Eq.~(\ref{localshearrate}). On the other hand, $\sigma_1$ is directly deduced from the torque $\Gamma(t)$ recorded by the rheometer [see Eq.~(\ref{sigma1})] and averaged over the same time window as that used for the velocity profile. The local shear stress $\sigma(r)$ is then computed from Eq.~(\ref{sigmar}). For each applied shear rate, the $\sigma(r)$ vs $\gp(r)$ data are reported in Fig.~\ref{rheol_3mm}. The experimental dispersion mainly arises from the estimation of the derivative in Eq.~(\ref{localshearrate}) which is based on a simple first-order approximation. This local flow curve is also compared to the global rheological data in Fig.~\ref{rheol_3mm}. In order to check that variations in the global flow behaviour remain small in spite of the very long durations of the start-up experiments, we first measured a flow curve $\sigma$ vs $\gp$ before starting the series of start-up experiments. Once start-up experiments were completed, another flow curve was recorded using the same protocol, namely a decreasing ramp of applied shear rate (from $100$~s$^{-1}$ down to $1.7\,10^{-2}$~s$^{-1}$) with a waiting time of 20~s per point. The data shown in Fig.~\ref{rheol_3mm} corresponds to the average of these two flow curves and the error bars indicate their difference. This difference reflects a global shift of the flow curve by a few pascals, which we attribute to a slow drift of the material properties due to repeated shearing protocols over more than $10^4$~s. The average flow curve is perfectly fit by a Herschel-Bulkley behaviour, $\sigma=\sigma_c+\tilde\eta\gp^n$, for shear rates larger than $3.5\,10^{-2}$~s$^{-1}$. At low shear rates, the deviation from Herschel-Bulkley behaviour (see inset of Fig.~\ref{rheol_3mm}) is attributed to paramount wall slip effects as already reported in the literature \cite{Meeker:2004a,Meeker:2004b}.

\begin{figure*}[!t]\tt
\centering
\includegraphics[width=0.85\linewidth]{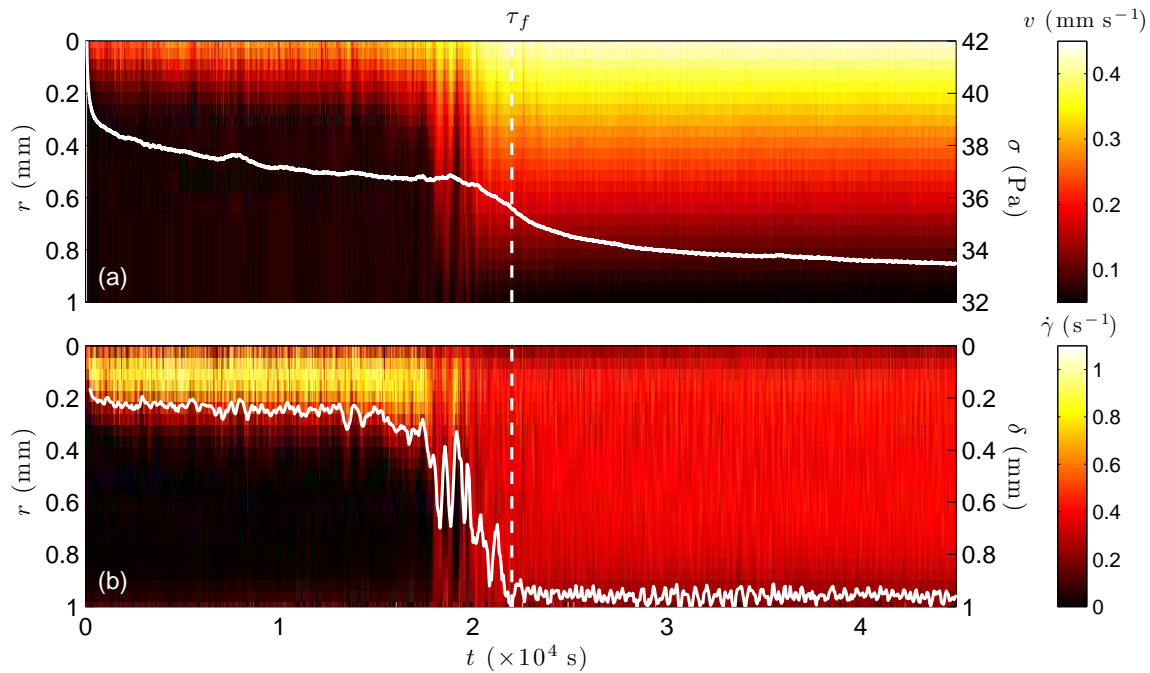}
\caption{(a) Spatiotemporal diagram of the velocity data $v(r,t)$ as a function of position $r$ and time $t$. A constant shear rate $\gp=0.5$~s$^{-1}$ is applied at time $t=0$ to a 1~\%~w/w carbopol microgel seeded with 0.5~\%~w/w hollow glass spheres in a smooth Couette cell of gap width 1~mm. The radial position $r$ (left vertical axis) is measured from the rotating inner wall. Also shown with a white line is the stress response $\sigma(t)$ (right vertical axis). (b) Spatiotemporal diagram of the local shear rate $\gp(r,t)$. The white line shows to the position $\delta(t)$ of the interface between the fluidized band and the solidlike region. The vertical dashed line indicates the fluidization time $\tau_f$, i.e. the time at which the shear rate field becomes homogeneous.}
\label{sptp_1mm_0p5}
\end{figure*}

The agreement between local and global rheology shown in Fig.~\ref{rheol_3mm} is quite remarkable. It can be further confirmed by computing the velocity profiles based on the Herschel-Bulkley behaviour inferred from the global rheological measurements and transferred locally:
\begin{equation}
\sigma(r)=\sigma_c+\tilde\eta\, \gp(r)^n \,\,\, \hbox{\rm for} \,\,\,\sigma(r) \ge \sigma_c\, .
\label{flowcurve2}
\end{equation}
Indeed, inserting Eq.~(\ref{flowcurve2}) into Eq.~(\ref{localshearrate}) and integrating over $r$ yields:
\begin{equation}
\frac{v(r)}{R_1+r} = \frac{v_1}{R_1}-\int_{0}^{r}{\,\frac{\hbox{\rm d}x}{R_1+x}\left[
\frac{\sigma_1\left(\frac{R_1}{R_1+x}\right)^2-\sigma_c}{\tilde\eta}\right]^{1/n}}\, ,
\label{vtheo2}
\end{equation}
where $v_1$ stands for the velocity of the fluid at the rotor, i.e. $v_1=v(0)$. In practice, $v_1$ is estimated by a linear extrapolation of the time-averaged velocity profile at $r=0$ and $\sigma_1$ is obtained from the rheometer measurement as explained above. Therefore, once the Herschel-Bulkley parameters $\sigma_c$, $\tilde\eta$, and $n$ are known from independent global measurements, the velocity profile can be predicted from Eq.~(\ref{vtheo2}) without any free parameter. The results of the computations based on Eq.~(\ref{vtheo2}) are superimposed to the experimental velocity profiles as red solid lines in Fig.~\ref{profiles_3mm}. Once again, the agreement is excellent, which shows that a single Herschel-Bulkley law allows to predict all the velocity profiles quite well. This confirms that our carbopol microgel behaves as a simple yield stress fluid in steady state, at least when confinement effects are negligible \cite{Goyon:2008,Ovarlez:2008,Goyon:2010}.

\subsection{General description of the fluidization process}

In this section, we describe qualitatively the fluidization process recorded after a shear rate $\gp=0.5$~s$^{-1}$ is applied at time $t=0$ to a carbopol sample in the smooth Couette cell of gap $e=1$~mm. Figure~\ref{sptp_1mm_0p5}(a) shows the spatiotemporal diagram where the velocity data $v(r,t)$ is coded in colour levels. Time $t$ and the radial position $r$ measured from the inner wall respectively correspond to the horizontal and to the vertical axis.  The white line in Fig.~\ref{sptp_1mm_0p5}(a) corresponds to the stress response $\sigma(t)$ recorded by the rheometer simultaneously to the local velocity and will be discussed in the next section.

The spatiotemporal diagram of Fig.~\ref{sptp_1mm_0p5}(b) shows the local shear rate $\gp(r,t)$ derived from the previous velocity data: $\gp(r,t)=r\partial (v(r,t)/r)\partial r$ where a second order differentiation scheme was used. To further reduce the noise level in the spatial derivative, $\gp(r,t)$ was smoothed using a moving average over four neighbouring points along the $r$-axis. From the spatiotemporal representation of $\gp(r,t)$, it is clear that the flow is spatially heterogeneous for $t\lesssim 2.10^4$~s: a strongly sheared region close to the rotor ($r<0.3$~mm) coexists with a solidlike region where the local shear rate vanishes ($r>0.3$~mm). This heterogeneous flow remains almost stationary for about 5 hours. At $t\sim 2.10^4$~s, the flow undergoes dramatic changes: strong fluctuations occur that lead to a completely homogeneous flow at larger times.

Therefore, although the microgel fluidization process starts with a long-lasting shear-banding regime, this heterogeneous flow does {\it not} correspond to the steady state which is rather characterized by the homogeneous shear rate field indeed expected for simple yield stress fluids. This fluidization scenario is entirely similar to that already described for a rough geometry~\cite{Divoux:2010}, which highlights its robustness with respect to boundary conditions.

 \subsection{Rheological signature of transient shear banding}

\begin{figure}[!t]
\centering
\includegraphics[width=0.9\columnwidth]{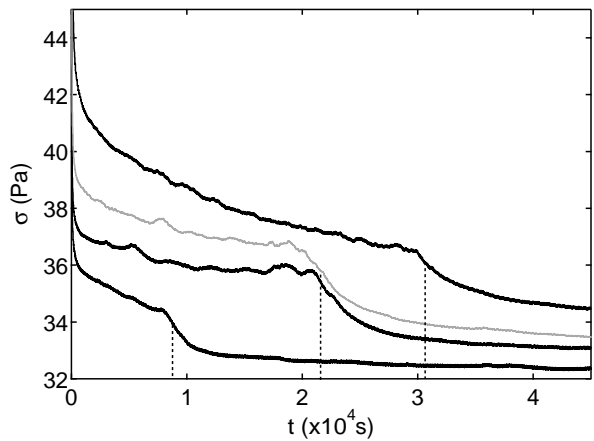}
\caption{Shear stress $\sigma$ as a function of time $t$ in a 1~\%~w/w ``pure'' carbopol microgel for different applied shear rates $\gp= 0.35$, 0.5, and 0.7~s$^{-1}$ (black lines, from top to bottom). The grey line is the stress response of the seeded microgel for $\dot \gamma= 0.5$~s$^{-1}$ already shown in Fig.~\ref{sptp_1mm_0p5}(a). Experiments performed in a smooth Couette cell of gap width 1~mm. The vertical dashed lines indicate the fluidization times as derived from the inflection points of $\sigma(t)$.}
\label{stress_seeding}
\end{figure}

As seen from the stress response $\sigma(t)$, superimposed to USV data as a white line in Fig.~\ref{sptp_1mm_0p5}(a), global rheological measurements strongly reflect the transient behaviour observed in simultaneous yet independent local velocity measurements. For $t\lesssim 2.10^4$~s, the shear stress slowly decays with noticeable fluctuations. Around $t\sim 2.10^4$~s, $\sigma(t)$ starts decreasing much more steeply before slowly leveling off to its steady-state value. Remarkably, the inflection point in the stress relaxation nicely corresponds to the time $\tau_f$ at which the flow becomes homogeneously sheared (see vertical dashed lines in Fig.~\ref{sptp_1mm_0p5}). The presence of such a strong rheological signature indicates that the fluidization process, observed with USV at about 15~mm from the cell bottom, probably occurs on a similar timescale throughout the whole height of the Couette cell. 

\begin{figure}[!t]
\centering
\includegraphics[width=0.9\columnwidth]{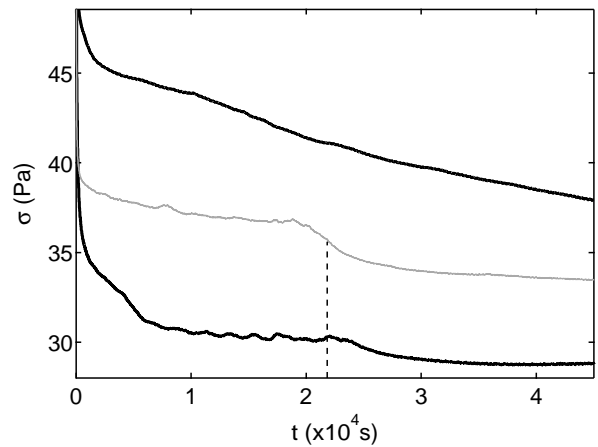}
\caption{Shear stress $\sigma$ as a function of time $t$ in two different geometries for similar shear rates: a rough plate-and-plate geometry (top black curve, $\gp=0.4$~s$^{-1}$) and a smooth cone-and-plate geometry (bottom black curve, $\gp=0.5$~s$^{-1}$). Experiments performed on the same batch of 1~\%~w/w carbopol microgels seeded with hollow glass spheres. This batch is different from that used previously in Fig.~\ref{sptp_1mm_0p5}. The grey line is the stress response for $\dot \gamma= 0.5$~s$^{-1}$ in a smooth Couette geometry of gap width 1~mm already shown in Fig.~\ref{sptp_1mm_0p5}(a)}
\label{stress_geom}
\end{figure}
\begin{figure*}[!t]\tt
\centering
\includegraphics[width=0.6\linewidth]{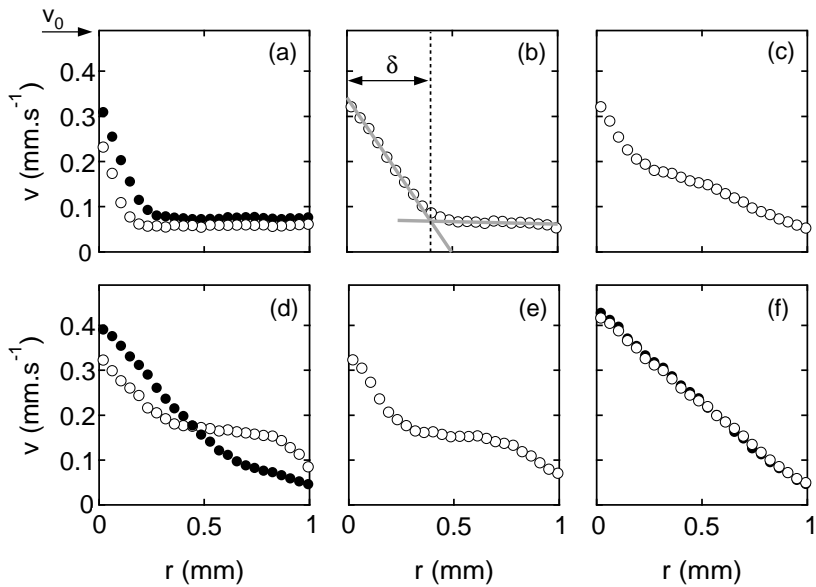}
\caption{Velocity profiles $v(r,t)$ extracted from the data shown in Fig.~\ref{sptp_1mm_0p5} for (a) $t=251$~s ($\circ$) and $t=7563$~s ($\bullet$), (b) $t=16720$~s, (c) $t=18021$~s, (d) $t=18671$~s ($\circ$) and $t=18871$~s ($\bullet$), (e) $19422$~s, and (f) $t=30981$~s ($\circ$) and $t=42742$~s ($\bullet$). The fitting procedure used to extract the slip velocities, the local shear rates in the two shear bands, and the position of the interface between the shear bands is illustrated by the grey lines in (b) (see text). Experiment performed in a smooth Couette cell of gap width 1~mm under an applied shear rate $\gp=0.5$~s$^{-1}$.}
\label{profiles_0p5}
\end{figure*}

Note that the sharp decrease of the stress at the shortest times is due to the presence of a stress overshoot that occurs for a strain $\gamma=\gp t \simeq 1$ but does not clearly show here due to the choice of vertical and horizontal scales. The stress overshoot phenomenon in the same carbopol microgels has been thoroughly investigated in Ref.~\cite{Divoux:2011b}. In particular, by focusing on the initial stage of the material response, we have shown that the stress first grows linearly with time. This is indicative of an elastic loading of the material, which is confirmed by the homogeneous strain field measured using USV. The stress overshoot then corresponds to a proliferation of plastic events that leads to the failure of the sample at the inner wall. This sudden failure gives way to total wall slip at the rotor. The shear band observed here then nucleates from the lubrication layer at the rotor. Full details about the stress overshoot and its correlation to the local flow behaviour can be found in Ref.~\cite{Divoux:2011b}.

Moreover, it is important to mention that, as seen in Fig.~\ref{stress_seeding} for various applied shear rates, the above rheological signature of transient shear banding is also observed in a ``pure'' carbopol sample, i.e. one that is free of seeding glass spheres. For $\gp= 0.5$~s$^{-1}$, the stress response is even quantitatively very close to that recorded at the same shear rate in the sample seeded with 0.5~\%~w/w hollow glass spheres (grey line in Fig.~\ref{stress_seeding}). Although no local velocimetry is available for the ``pure'' sample, the very similar rheological response is surely indicative of the same fluidization process. We infer from Fig.~\ref{stress_seeding} that our contrast agents have no significant influence on the transient shear banding phenomenon.

We also check in Fig.~\ref{stress_geom} that such a striking feature is also seen in the cone-and-plate geometry. In that case, the ``kink'' in $\sigma(t)$ for $\gp= 0.5$~s$^{-1}$ again coincides with that observed in the Couette geometry, in spite of a shift in the absolute value of the stress which may be attributed to the fact that these experiments were performed on different microgel batches. However, in the plate-and-plate, a kink followed by an inflection point is far more difficult to detect: in this case, the fluidization time could be virtually anywhere above $10^4$~s. This is most probably due to the large shear rate heterogeneity in the plate-and-plate geometry, where the local shear rate varies from zero at the rotation axis to the applied value $\gp$ at the plate periphery. We shall see in Sect.~\ref{s.shearrate} that the fluidization time increases dramatically with decreasing shear rates. Therefore, the fluidization process in a plate-and-plate geometry could be seen as the superposition of fluidization processes with widely different timescales, leading to a stress response that does not show very sharp features. Finally, it is worth noticing that the surface roughness of the shearing tools does not seem to play any prominent role since the exact same shape has also been reported for $\sigma(t)$ in a Couette cell of surface roughness 60~$\mu$m.\cite{Divoux:2010}

To conclude, the transient shear banding phenomenon and the subsequent fluidization have a characteristic rheological signature that appears very robust with respect to the shearing geometry and to the boundary conditions, at least for the low shear rates reported so far ($\gp=0.3$--0.7~s$^{-1}$). In the following, we shall show that the kink in the stress response actually disappears when the shear rate or the gap width is increased, although transient shear banding is still observed through USV.

\subsection{Quantitative analysis of the velocity profiles}
\label{s.profiles}

\begin{figure}[!t]\tt
\centering
\includegraphics[width=0.9\columnwidth]{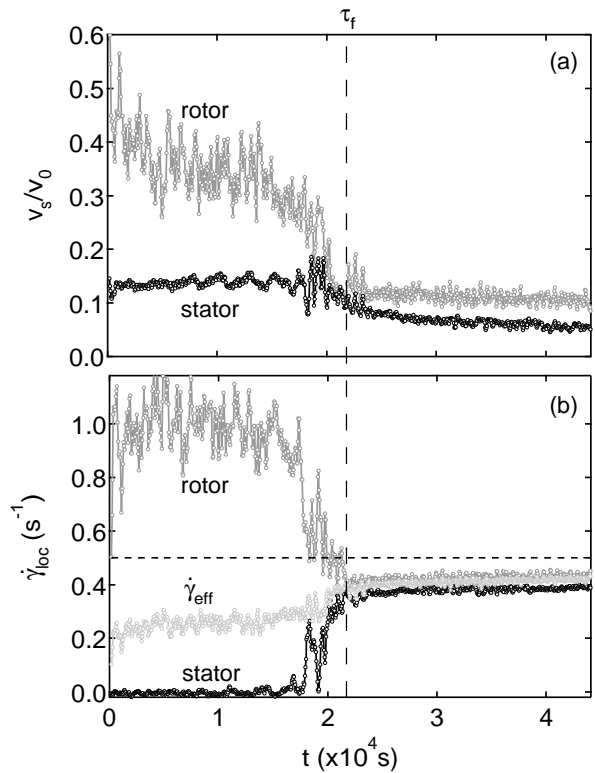}
\caption{Analysis of the velocity data shown in Fig.~\ref{sptp_1mm_0p5}. (a) Slip velocities $v_s$ at the stator (black) and at the rotor (grey). (b) Local shear rates $\gloc$ at the stator (black) and at the rotor (grey) together with the effective shear rate $\geff$ (light grey). The vertical dashed line indicates the fluidization time $\tau_f$. The horizontal dotted line in (b) shows the shear rate applied by the rheometer $\gp=0.5$~s$^{-1}$. Experiment performed in a smooth Couette cell of gap width 1~mm.}
\label{analysis_0p5}
\end{figure}

Velocity profiles typical of the various stages of the fluidization process were extracted from the data in Fig.~\ref{sptp_1mm_0p5}(a) and plotted in Fig.~\ref{profiles_0p5}: quasi-stationary shear-banded velocity profiles for $t< 1.7\,10^4$~s [Fig.~\ref{profiles_0p5}(a) and (b)], strongly fluctuating profiles for $t\sim1.7$--2.$10^4$~s [Fig.~\ref{profiles_0p5}(c--e)], and fully fluidized, homogeneous and stationary profiles for $t> 2.10^4$~s [Fig.~\ref{profiles_0p5}(f)]. A movie of the velocity profiles is also available as supplementary material.$\dag$ During the abrupt fluidization at $t\sim 2.10^4$~s, velocity profiles with three bands are observed transiently as in Fig.~\ref{profiles_0p5}(d) and (e). Such strong fluctuations and anomalous velocity profiles are reminiscent of those observed in laponite suspensions sheared in a smooth geometry.\cite{Gibaud:2008,Gibaud:2009} Here however, no periodic oscillations between banded and linear velocity profiles are observed and these erratic fluctuations are observed over a rather limited time period compared to the total duration of the transient regime.

\begin{figure}[!t]\tt
\centering
\includegraphics[width=0.9\columnwidth]{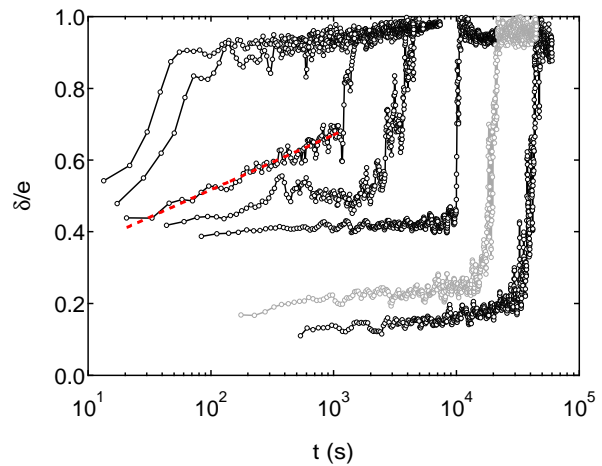}
\caption{Width $\delta$ of the shear band normalized by the gap width $e$ versus time for different applied shear rates $\gp=0.35$, 0.5, 0.9, 1.5, 2.0, 2.3, and 3.0~s$^{-1}$ from right to left. The fluidization time strongly decreases with $\gp$. The gray line corresponds to the shear rate $\gp=0.5$~s$^{-1}$ shown in Fig.~\ref{sptp_1mm_0p5}. The red dashed line is a logarithmic fit for $\gp=2.0$~s$^{-1}$ and $t<10^3$~s. Experiments performed in a smooth Couette cell of gap width 1~mm.}
\label{delta_1mm}
\end{figure}

Important quantitative information can be readily extracted from individual velocity profiles. As shown in Fig.~\ref{profiles_0p5}(b), linear fits in both shear bands directly yield the local shear rates $\gloc$ in the flowing and solidlike regions respectively. Extrapolating these fits to $r=0$ and $r=e$ allows one to estimate the microgel velocities $v(r=0)$ and $v(r=e)$ at the vicinity of the shearing surfaces, and therefore the slip velocities $v_s$ at both walls, as well as the effective shear rate $\geff=[v(r=0)-v(r=e)]/e$. Finally, the intersection of the two linear fits yields the position $\delta$ of the interface between the fluidized and solidlike regions. In the data presented in Fig.~\ref{analysis_0p5}, an additional moving average over four consecutive data points in time is applied to $v_s(t)$, $\gloc(t)$, $\geff(t)$, and $\delta(t)$ in order to smooth out noise on short timescales.

It can be checked that our determination of $\delta$, plotted both in Fig.~\ref{sptp_1mm_0p5}(b) as a white line and in Fig.~\ref{delta_1mm} as a grey line, indeed corresponds to the transition from high local shear rate to zero shear in the transient shear-banding regime. The fluidization time $\tau_f$ is then defined as the time at which the shear band totally disappears, i.e. $\delta(\tau_f)\simeq e$. The fact that $\delta$ never reaches exactly $e$ is due to our procedure based on the intersection of two fits that use at least two data points close to the walls. Still, $\tau_f$ is always a well-defined quantity as shown by the dashed line in Fig.~\ref{sptp_1mm_0p5}(b).

\begin{figure*}[!t]\tt
\centering
\includegraphics[width=0.85\linewidth]{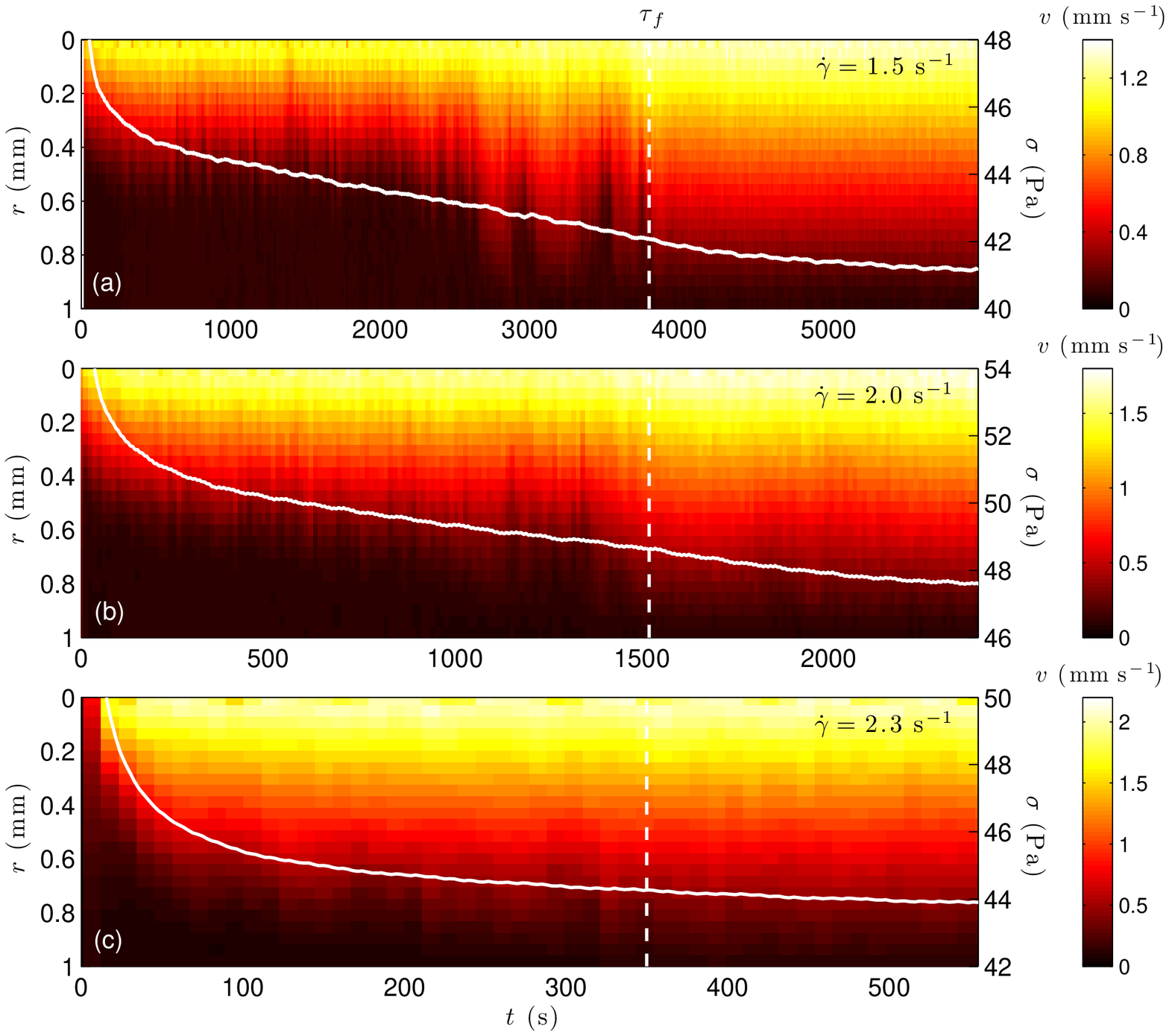}
\caption{Spatiotemporal diagrams of the velocity data $v(r,t)$ for an applied shear rate of (a)~$\gp=1.5$~s$^{-1}$, (b)~$\gp=2.0$~s$^{-1}$, and (c)~$\gp=2.3$~s$^{-1}$. White lines are the corresponding stress responses $\sigma(t)$ (right vertical axis). The vertical dashed lines indicate the fluidization times $\tau_f$. The time interval between two velocity profiles is 17~s, 13~s, and 11~s in (a), (b), and (c) respectively. Experiments performed in a smooth Couette cell of gap width 1~mm.}
\label{sptp_1mm_1p5_2_2p3}
\end{figure*}

Figure~\ref{analysis_0p5} shows the results of the analysis of the velocity data of Fig.~\ref{sptp_1mm_0p5}(a). An important amount of wall slip is observed throughout the shear banding regime with slip velocities up to 40~\% of the rotor imposed velocity $v_0$ at the rotating inner wall and up to 15~\% of $v_0$ at the fixed outer wall [see Fig.~\ref{analysis_0p5}(a)]. Both slip velocities drop by about a factor of two as the shear band disappears and the flow becomes homogeneous around $\tau_f$. In steady state, the slip velocities relative to $v_0$ are about 10~\% at the rotor and 5~\% at the stator. This should be contrasted to our previous results in a rough Couette cell where the steady state did not show any significant wall slip.\cite{Divoux:2010}

The above evolution of the slip velocities is reflected in the effective shear rate: $\geff$ first increases very slowly from about 0.2 to 0.3~s$^{-1}$ in the shear-banding regime and then more rapidly up to roughly 0.4~s$^{-1}$, which remains significantly below the imposed value of 0.5~s$^{-1}$ due to the presence of wall slip in steady state [see Fig.~\ref{analysis_0p5}(b)]. As expected from the velocity profiles, the local shear rate close to the stator remains zero until the flow field shows strong fluctuations for $t\sim 2.10^4$~s, while the local shear rate in the flowing shear band is about 1~s$^{-1}$ in the shear-banding regime and sharply falls down to 0.4~s$^{-1}$ during full fluidization. In the homogeneous flow regime ($t>\tau_f$), $\geff$ and both $\gloc$ collapse to within experimental precision.

Finally, it is worth emphasizing again two important observations drawn from the above analysis. First, as already noted for the stress response $\sigma(t)$ in Fig.~\ref{sptp_1mm_0p5}(a), noticeable fluctuations are reported for slip velocities and local shear rates during transient shear banding whereas the measurements become much smoother once full fluidization is achieved. This points to heterogeneous spatiotemporal dynamics and a two-dimensional view of the local flow field would certainly help to clarify the origin of these fluctuations. Second, our results highlight the importance of very long start-up experiments to ensure that a steady state is reached, especially close to the yield stress, i.e. at small imposed shear rates. Indeed, if the experiment shown above had been stopped after 100~s, 1000~s, or even $10^4$~s, one could have mistaken the shear-banded state for the steady state. Here, a total duration of more than $4.10^4$~s allows one to reach the true steady state, namely the homogeneous flow characteristic of a simple yield stress fluid.

\begin{figure}[!t]\tt
\centering
\includegraphics[width=0.9\columnwidth]{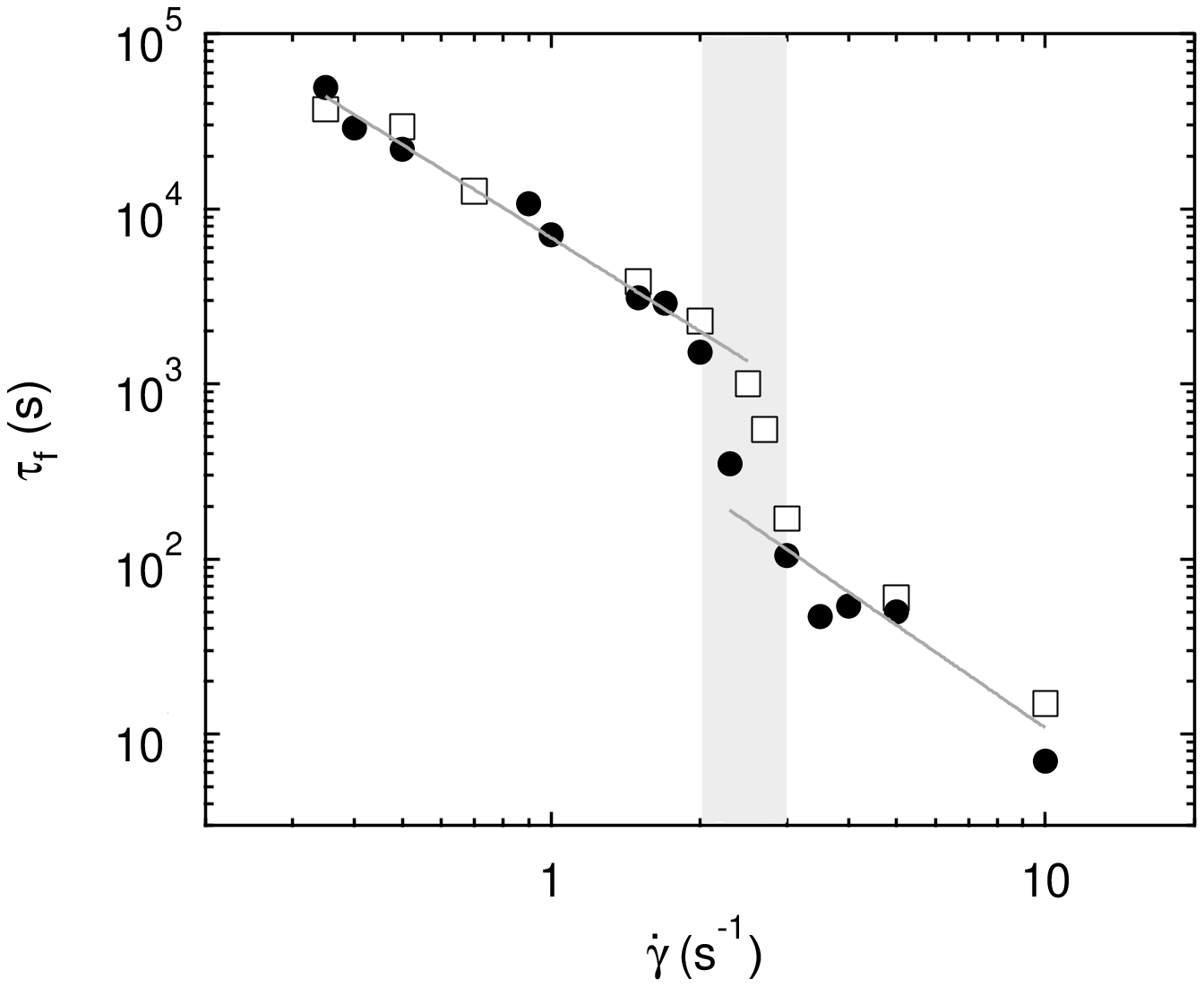}
\caption{Fluidization time $\tau_f$ as a function of the applied shear rate $\gp$ for two different batches of 1~\%~w/w carbopol microgels. The grey solid lines are the best power-law fits $\tau_f=A/\gp^\alpha$ of the data for $\gp\le 2$~s$^{-1}$ ($A=7.0\,10^3$~s$^{1-\alpha}$ and $\alpha=1.8\pm 0.1$) and for $\gp\ge 3$~s$^{-1}$ ($A=960$~s$^{1-\alpha}$ and $\alpha=1.9\pm 0.3$). The shaded area indicates $\gps\simeq 2.5\pm 0.5$~s$^{-1}$ that separates a transient shear-banding regime characterized by an induction period followed by strong fluctuations from a fast and smooth shear-banding regime where the shear band grows continuously (see text). Filled symbols correspond to the carbopol batch investigated so far in Fig.~\ref{sptp_1mm_0p5} and Figs.~\ref{profiles_0p5}--\ref{sptp_1mm_1p5_2_2p3}. Experiments performed in a smooth Couette cell of gap width 1~mm.}
\label{tauf_1mm}
\end{figure}

\subsection{Evolution of the fluidization process with the shear rate}
\label{s.shearrate}

To test the robustness of the fluidization scenario described in the previous sections, the imposed shear rate $\gp$ was systematically varied from 0.35 to 10~s$^{-1}$. In all cases, transient shear banding was observed and the fluidization time was found to decrease sharply with $\gp$. Figure~\ref{delta_1mm} gathers some measurements of the width $\delta(t)$ of the shear band as a function of time. It is clear that the fluidization proceeds faster as $\gp$ is increased. Moreover, for small shear rates, typically $\gp\lesssim1$~s$^{-1}$, the shear band remains almost stationary so that the transient shear banding resembles an ``induction'' period after which the fluidization process dramatically accelerates. For intermediate shear rates, $1\lesssim\gp\lesssim 2$~s$^{-1}$, the non-zero initial slope of $\delta(t)$ indicates that the shear band slowly grows during the transient shear-banding regime. At $\gp=2$~s$^{-1}$, a logarithmic growth is observed for $t<10^3$~s (see red dashed line in Fig.~\ref{delta_1mm}), before a sudden acceleration occurs similar to that reported at lower shear rates. 

Yet, for $\gp>2$~s$^{-1}$, there is no such a sharp change that separates the shear-banding regime from the homogeneous flow regime: the shear band rather grows continuously across the gap of the Couette cell. Moreover, between $\gp=2$~s$^{-1}$ and $\gp=2.3$~s$^{-1}$, there seems to be a very large drop in the fluidization time for an increase in the applied shear rate of only 15~\%. Therefore, we believe that the fluidization scenario goes from an induction-like process followed by an abrupt acceleration with large fluctuations at low $\gp$ to a much smoother and faster process with no induction time at high $\gp$.

Figure~\ref{sptp_1mm_1p5_2_2p3} further illustrates this change of behaviour by comparing the spatiotemporal diagrams of $v(x,t)$ measured at $\gp=1.5$~s$^{-1}$, $\gp=2$~s$^{-1}$, and $\gp=2.3$~s$^{-1}$ (see also the movies in the supplementary material\dag). The horizontal scale was chosen so that the fluidization times, respectively 3800~s, 1520~s, and 350~s, graphically coincide. Although the temporal resolution is not as good as for $\gp=1.5$~s$^{-1}$ due to the smaller time scale, the shear band cannot be seen to take any quasi-stationary position for $\gp=2.3$~s$^{-1}$ as is the case for $t\lesssim 2500$~s at $\gp=1.5$~s$^{-1}$. Fluctuations for $t\lesssim \tau_f$ also seem to be absent for the larger shear rate. Finally, the stress signal for $\gp=2.3$~s$^{-1}$ does not show any sign of kink whereas $\sigma(t)$ does display a small but detectable inflection point around $\tau_f$ for $\gp=1.5$~s$^{-1}$. The experiment at $\gp=2$~s$^{-1}$ appears as an intermediate case where fluctuations close to $\tau_f$ become negligible and the kink in $\sigma(t)$ is hardly visible yet the shear band grows very slowly around $\delta\simeq 0.6$~mm and suddenly accelerates around $\tau_f$. To us, these observations imply that the nature of the transient shear-banding regime changes between the two shear rates shown in Fig.~\ref{sptp_1mm_1p5_2_2p3}(a) and (c), i.e. for a characteristic shear rate $\gps\simeq 2.0$--2.3~s$^{-1}$.

The evolution of $\tau_f$ as a function of $\gp$ is shown in Fig.~\ref{tauf_1mm}. The measurements performed on the batch investigated so far are plotted in filled circles together with another data set obtained on a different batch in the same smooth Couette geometry (open squares). In this case, reproducibility is very good except maybe at the change of shear-banding regime around $\gps$ (see shaded area). In Ref.~\cite{Divoux:2010}, we reported that the fluidization time decreases as a power-law of $\gp$. In the present data, however, the change of banding regime at $\gps$ shows up as a step down in $\tau_f$. Still, the fluidization time behaves as a power law if one considers $\gp<\gps$ and $\gp>\gps$ separately. Moreover, the best fits yield similar exponents of 1.8 and 2.0 for the two regimes so that the change in shear-banding only appears as a change in the prefactor. We shall further discuss this power-law dependence in Section~\ref{discuss}.

\subsection{Effects of the gap width and of boundary conditions}

\begin{figure}[!t]\tt
\centering
\includegraphics[width=0.9\columnwidth]{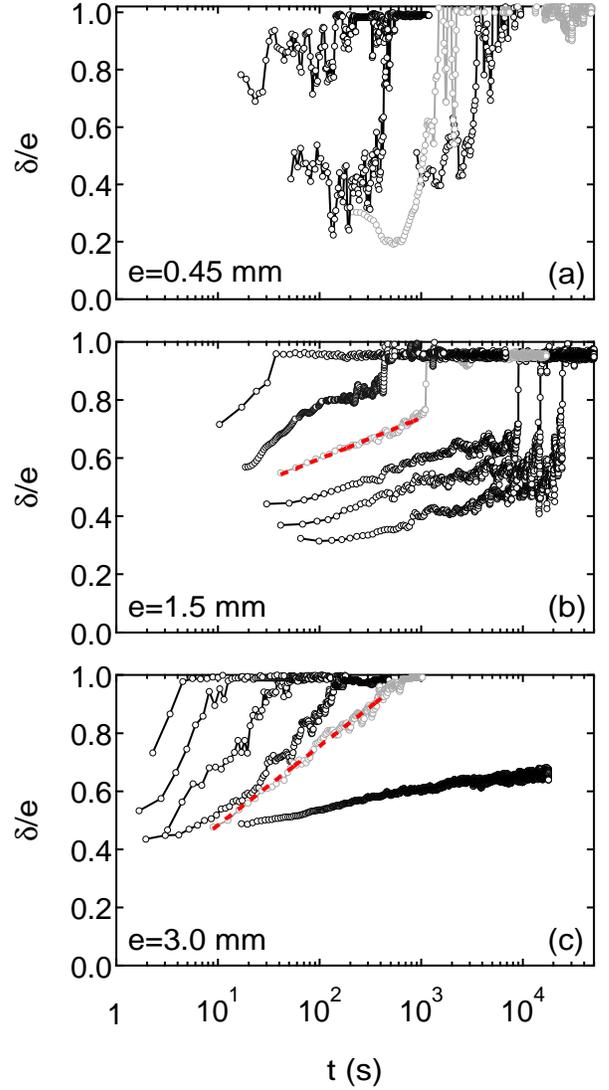}
\caption{Width $\delta$ of the shear band normalized by the gap width $e$ versus time for different gap widths $e$ and applied shear rates $\gp$.
(a)~$e=0.45$~mm and $\gp=0.6$, 0.85, 1.5, and 3~s$^{-1}$ from right to left.
(b)~$e=1.5$~mm and $\gp=0.4$, 0.5, 0.65, 0.8, 1.5, and 2~s$^{-1}$ from right to left.
(c)~$e=3$~mm and $\gp=0.5$, 0.7, 1.2, 2, 3, and 4.5~s$^{-1}$ from right to left.
Gray lines correspond to the shear rates shown in Supplementary Figure~2 and red dashed lines are logarithmic fits of $\delta(t)$. Experiments performed in smooth Couette cells on different batches of 1~\%~w/w carbopol microgels.}
\label{delta_gaps}
\end{figure}
\begin{figure}[!t]\tt
\centering
\includegraphics[width=0.9\columnwidth]{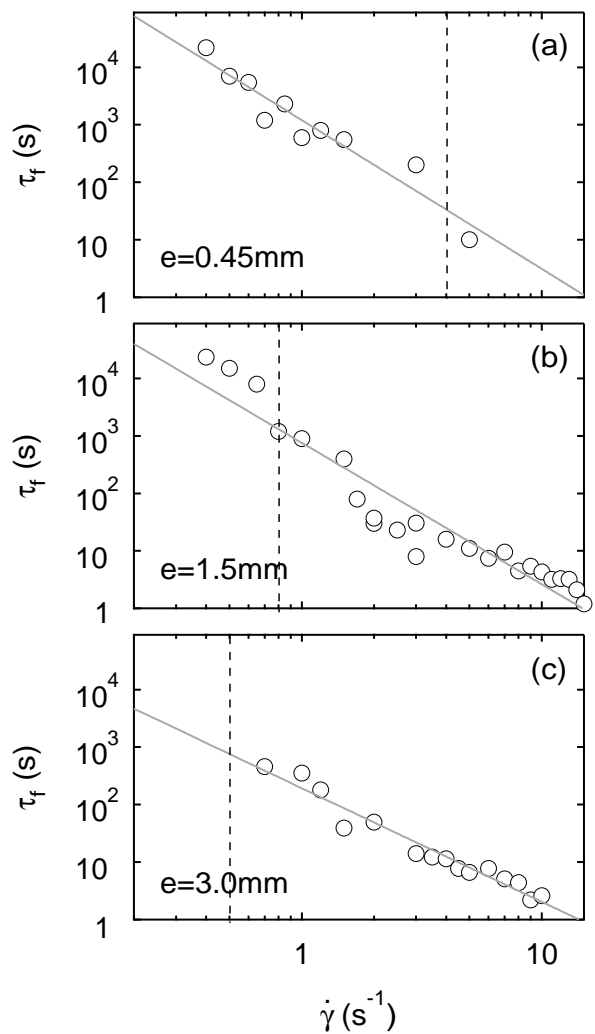}
\caption{Fluidization time $\tau_f$ as a function of the applied shear rate $\gp$ for various gap widths. 
The solid lines are the best power-law fits of the data $\tau_f=A/\gp^\alpha$.
(a) $e=0.45$~mm: $A=1210$ and $\alpha=2.6\pm 0.2$.
(b) $e=1.5$~mm: $A=759$ and $\alpha=2.4\pm 0.2$.
(c) $e=3$~mm: $A=192$ and $\alpha=2.0\pm 0.1$.
The vertical dashed line indicates $\gps$ (see text). Experiments performed in smooth Couette cells on different batches of 1~\%~w/w carbopol microgels.}
\label{tauf_gaps}
\end{figure}

In this section, we investigate the effects of the shearing geometry by first varying the gap width from 0.45 to 3~mm and then turning to rough walls. In all cases, the steady state is characterized by homogeneous velocity profiles (with an amount of wall slip that depends on the gap width for smooth geometries, see Supplementary Figure~1). Moreover, a transient shear-banding scenario similar to that described above is always found and a fluidization time $\tau_f$ can be defined as explained in Sect.~\ref{s.profiles}.

\subsubsection{Influence of the gap width.~}
\label{s.gap}

Figure~\ref{delta_gaps} shows measurements of $\delta(t)$ similar to those of Fig.~\ref{delta_1mm} but performed in smooth Couette cells of gap widths $e=0.45$, 1.5, and 3~mm. While the shear-banding regime is characterized by large fluctuations of $\delta(t)$ over almost the whole range of investigated shear rates for the smallest gap width [see Fig.~\ref{delta_gaps}(a)], shear banding proceeds in a smooth, continuous manner for most of the shear rates under study at the largest gap width [see Fig.~\ref{delta_gaps}(c)]. Therefore, there exists a clear dependence of the transient shear-banding regime on the gap width: the characteristic shear rate $\gps$ that separates quasi-stationary then intermittent banding from smooth and continuous banding decreases with $e$. We determined $\gps\simeq 4$, 0.8, and 0.5~s$^{-1}$ for $e=0.45$, 1.5, and 3~mm respectively. These estimates are based on a close inspection of the spatiotemporal velocity data: we take the existence of (i) a quasi-stationary phase, (ii) strong velocity fluctuations, or (iii) an abrupt jump in $\delta(t)$ before complete fluidization as evidences that  $\gp<\gps$. We also emphasize that for the larger gaps, the evolution of $\delta(t)$ close to or above $\gps$ is well fitted by a logarithmic growth (see red dashed lines in Fig.~\ref{delta_gaps}). In short, increasing the gap width at a given applied shear rate has the same qualitative effect as increasing the shear rate for a given gap width (see also Supplementary Figure~2 and the corresponding movies in the supplementary material\dag.)

Finally, the fluidization times measured for $e=0.45$, 1.5, and 3~mm are shown in Fig.~\ref{tauf_gaps}. The characteristic shear rates $\gps$ discussed above are indicated as dashed lines. For the lowest shear rate investigated at $e=3$~mm ($\gp=0.5$~s$^{-1}$), quasi-stationary banding was observed but the experiment duration was not enough to allow for a measurement of $\tau_f$. Therefore, the corresponding point is not reported in Fig.~\ref{tauf_gaps}(c) but this experiment still allowed for an estimate of $\gps$. We shall further discuss Fig.~\ref{tauf_gaps} below in Sect.~\ref{s.powlaw}, together with results obtained using different boundary conditions.

\begin{figure}[!t]\tt
\centering
\includegraphics[width=0.9\columnwidth]{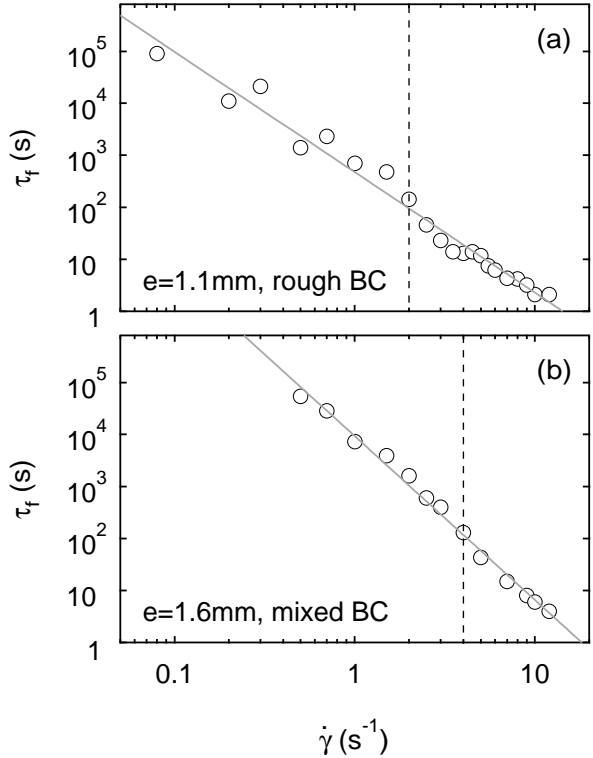}
\caption{Fluidization time $\tau_f$ as a function of the applied shear rate $\gp$ for various boundary conditions. 
The solid lines are the best power-law fits of the data $\tau_f=A/\gp^\alpha$.
(a) Rough Couette cell of gap $e=1.1$~mm: $A=472$ and $\alpha=2.3\pm 0.1$.
(b) Couette cell of gap $e=1.6$~mm with a rough rotor and a smooth stator: $A=9228$ and $\alpha=3.1\pm 0.1$.
The vertical dashed line indicates $\gps$. Experiments performed on different batches of 1~\%~w/w carbopol microgels.}
\label{tauf_bc}
\end{figure}

\subsubsection{Influence of boundary conditions.~}
\label{s.bc}

The possible influence of boundary conditions, i.e. surface roughness and physico-chemistry of the cell walls, is another important issue to be explored. In our previous works,\cite{Divoux:2010,Divoux:2011a} we have shown that using sand paper instead of polished Plexiglas as shearing surfaces does not affect the existence of transient shear banding. Except for early stages immediately following the stress overshoot,\cite{Divoux:2011a} the shear band develops in the same manner for both boundary conditions. As long as the same carbopol batches are considered, the fluidization times obtained with different boundary conditions are even quantitatively close to each other as shown in Fig.~4 of Ref.~\cite{Divoux:2010}. 

The data for rough boundary conditions was replotted in Fig.~\ref{tauf_bc} together with the fluidization times obtained on a different carbopol batch under mixed boundary conditions (rough rotor, smooth stator, $e=1.6$~mm, see Sect.~\ref{s.rheomes}). For the experiments performed in the rough Couette cell of gap 1.1~mm, we estimated $\gps\simeq 2$~s$^{-1}$, close to the case of a smooth Couette cell of gap 1~mm (see Sect.~\ref{s.shearrate}), while one gets $\gps\simeq 3$~s$^{-1}$ for mixed boundary conditions and $e=1.6$~mm, significantly above the value found for the smooth geometry of gap $e=1.5$~mm. 

Therefore, although it is clear that the most salient features of transient shear banding are not affected by boundary conditions, subtle effects on $\gps$ may be induced by surface effects. Further investigations using systematically the same carbopol batch are still required to settle this open issue.

\subsubsection{Robustness of the power-law behaviour for $\tau_f$ vs $\gp$.~}
\label{s.powlaw}

Contrary to what is seen in Fig.~\ref{tauf_1mm} for $e=1$~mm, the fluidization times in Figs.~\ref{tauf_gaps} and \ref{tauf_bc} do not display any sharp break around $\gps$. This may be due to a lack of data for $\gp\simeq\gps$ but also to a complex dependence of the drop in $\tau_f$ on the gap width or on the boundary conditions. In any case, we chose to fit the $\tau_f$ data of Figs.~\ref{tauf_gaps} and \ref{tauf_bc} by a single power law over the whole range of investigated shear rates. We find that the power-law behaviour always provide a good description of $\tau_f$ vs $\gp$ with an exponent $\alpha=2.0$--3.1 that may depend on the carbopol batch as already noted in Refs.~\cite{Divoux:2010,Divoux:2011a}. Consequently, the power-law behaviour of $\tau_f$ vs $\gp$ appears as very robust with respect to a change in gap width or boundary conditions.

\section{Discussion and outlook}
\label{discuss}

\subsection{Towards a ``phase diagram'' for transient shear banding}
\label{s.phasediag}

\begin{figure}[!t]\tt
\centering
\includegraphics[width=0.9\columnwidth]{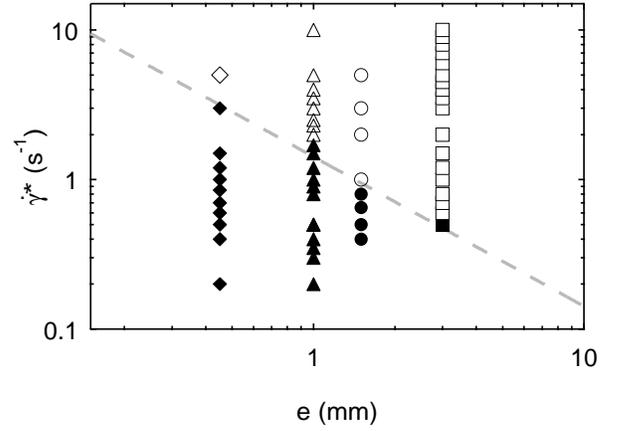}
\caption{Type of transient shear-banding regime as a function of the gap width $e$ and the applied shear rate $\gp$. Filled symbols indicate a transient shear banding characterized by an induction period followed by strong fluctuations. Open symbols correspond to a fast and smooth shear-banding regime. The dashed line is $\gps = 1.5/e$. Experiments performed in smooth Couette geometries for different batches of 1~\%~w/w carbopol microgels.}
\label{phasediag}
\end{figure}

From Sect.~\ref{s.shearrate} and \ref{s.gap} above, one concludes that the gap width and the shear rate are two parameters that control the way transient shear banding proceeds. This leads us to summarize all our data in a single ``phase diagram'' in the $(e,\gp)$ plane as shown in Fig.~\ref{phasediag}. Let us recall that at low values of $\gp$ or $e$, the transient shear band remains almost stationary for several hours before strong fluctuations lead to homogeneous flow, whereas for larger values of $\gp$ and $e$ the shear band invades the whole gap in a much smoother way. A rough estimate for the boundary between the two different shear-banding regimes is $\gps\sim 1/e$ (see dashed line in fig.~\ref{phasediag}).

First, such a scaling suggests the existence of some underlying critical velocity $v^\star\simeq \gps e$ strongly reminiscent of the ``slip velocity'' introduced in Refs.~\cite{Meeker:2004a,Meeker:2004b}. In this framework, one could speculate that for an imposed value of the rotor velocity $v_0<v^\star$, wall slip triggers an erosion process from the wall toward the bulk material. This process fragilizes the gel and results in its abrupt fluidization (see also Sect.~\ref{s.mechanism} below). For large rotor velocities, $v_0>v^\star$, the fluidization scenario is rather homogeneous in the bulk and controlled by $v^\star$ and thus by the gel properties. Such an interpretation remains speculative and deserves to be tested by with a systematic control of the microscopic gel properties.

Second, another possible interpretation of the effect of the gap width lies in the stress heterogeneity inherent to the Couette geometry. Indeed, between the two concentric cylinders, the stress decreases as $1/(R_1+r)^2$, where $R_1$ is the rotor radius. In other words, the ratio of the stress $\sigma_1$ at the rotor to the stress $\sigma_2$ at the stator is $\sigma_1/\sigma_2=(R_2/R_1)^2$, where $R_2$ is the stator radius. When the gap width is increased from 0.45~mm to 3~mm, the stress heterogeneity increases from 4~\% to about 25~\%. Therefore one could conclude that this stress heterogeneity somehow promotes the smooth and continuous transient shear banding regime. However, our observations could also be due to confinement rather than to stress heterogeneity. At this stage, more experiments are required to discriminate between the effect of stress confinement and that of the curved geometry. 

\subsection{A possible microscopic interpretation of the fluidization scenario}
\label{s.mechanism}

In this subsection, we provide the reader with a possible microscopic scenario of the fluidization process in agreement with our macroscopic observations. In particular, we discuss an hypothetic interpretation of the puzzling logarithmic growth of $\delta (t)$ before the sudden fluidization, which is reminiscent of structural aging phenomena widely observed in other soft glassy materials under stress, such as granular materials,\cite{Bocquet:1998,Gayvallet:2002} gelatin gels,\cite{Ronsin:2009} or weakly attractive suspensions.\cite{Barlett:2011}

This scenario goes as follows: carbopol microgels are composed of soft jammed domains, a few microns in diameter, reversibly linked to one another by polymers and thus forming a soft jammed structure. Upon starting a step-rate experiment, the stress first builds up and the gel experiences linear deformation after which plastic events take place in material.\cite{Divoux:2011b} One can imagine that plasticity occurs when soft domains sharing some polymers are being disconnected. Then, the stress reaches a maximum before decreasing. Concomitantly to this stress overshoot, the gel breaks at the rotor leading to a thin highly sheared lubrication layer which is smaller than 40 microns, the spatial resolution of USV. In our previous work,\cite{Divoux:2011b} we estimated the viscosity of the lubrication layer to be close to that of water so that we can assume that it contains almost no carbopol. Next, the gel enters the so-called ``induction phase'' during which the gel is slowly eroded from the rotor before suddenly fluidizing at a well defined time $\tau_f$.

We propose that the erosion process during the induction phase somehow fragilizes the bulk arrested microgel bringing it to a {\it critical state} before complete sudden fluidization occurs. Such a critical state could be analog to the one reached by a colloidal gel experiencing ``delayed sedimentation,'' right before its collapse (see for instance Fig.~8(b) in Ref.~\cite{Barlett:2011} which is strongly reminiscent of Fig.~\ref{delta_1mm} in the present manuscript). Let us emphasize that the delayed phenomenon reported in Ref.~\cite{Barlett:2011} is linked to a thermally activated process in the absence of any macrocopic motion. Similarly, during the induction phase, the arrested gel undergoes solid-body rotation, i.e. no macroscopic shear. Note however that such a scenario is certainly not generic of all the delayed sedimentation processes reported in the literature. Indeed, some systems involve fractures, channel formation, and convection\cite{Poon:1999,Manley:2005} which are not seen in the present delayed fluidization process. The comparison with delayed sedimentation prompts us to go one step further and compare the fluidization process of our microgel to a solid-liquid phase transition at low temperature.\cite{Testard:2011} In the next subsection, we dwell on such an analogy in terms of an out-of-equilibrium phase transition and critical phenomena.

How exactly the arrested gel gets fragilized into a critical state remains an open question. One can think of a velocity-controlled process that involves slip effects as discussed in the previous subsection. One could also argue in favor of a stress-controlled process in which the viscous drag from the highly-sheared lubrication layer and from the already fluidized material could play the role of a small external drive on the arrested microgel. Indeed, due to this viscous drag, the arrested gel is likely to be submitted to some shear stress below the yield stress of the microgel. Such a small stress perturbation could be enough to induce creep behaviours and the aging of the microgel in the non-flowing band as already oberved for different microgels \cite{Cloitre:2000,Purnomo:2008} as well as other yield stress fluids \cite{Coussot:2006,Joshi:2008b}.

\subsection{A tentative interpretation in terms of critical phenomena}
\label{s.critical}

In this subsection, we propose an interpretation of the  power-law behaviour of the fluidization time in terms of critical phenomena based on an analogy between flows close to the yield stress and thermodynamic transitions close to a critical temperature. This analogy is suggested by the power-law dependence of the fluidization times as a function of the distance to the yield point. Indeed, we have shown that, under  controlled stress, the fluidization time depends on the difference between the imposed stress $\sigma$ and the yield stress $\sigma_c$ as:\cite{Divoux:2011a} $\tau_f^{(\sigma)}\sim (\sigma-\sigma_c)^{-\beta}$. Under controlled shear-rate, the fluidization time depends on $\dot\gamma$ as:\cite{Divoux:2010} $\tau_f^{(\dot\gamma)}\sim \dot\gamma^{-\alpha}$. Moreover,  assuming that the fluidization processes under controlled strain or stress are governed by the same physical mechanisms, i.e. $\tau_f^{(\sigma)}\sim\tau_f^{(\dot\gamma)}$, one recovers the Herschel-Bulkley rheology $\delta\sigma=\sigma-\sigma_c=A\dot\gamma^{n}$ with the exponent $n=\alpha /\beta$. It is important to note that the exponent $n$\cite{Roberts:2001,Baudonnet:2004,Lee:2011} and both fluidization exponents $\alpha$ and $\beta$ depend on the microgel concentration and on the microgel preparation, but that the ratio $\alpha/\beta$ remains equal to the Herschel-Bulkley exponent $n$ whatever the microgel batch.

Here, we propose scaling arguments to describe the power-law dependences of the fluidization time. 
If we consider the situation in which a controlled shear rate $\dot\gamma$ is imposed at constant particle density, then simulations of particle diffusion\cite{Lemaitre:2009} and of diffusion and flow heterogeneity\cite{Martens:2011} suggest that one should think of the limit $\dot\gamma\to 0$ as a critical point in the space of shear driven steady states, characterized by a diverging time scale.  Regarding the viscous part of the shear stress $\delta\sigma$ as an order parameter conjugate to the ``field" $\dot\gamma$, we therefore propose a critical scaling law  \cite{Barton:1975} for {\it steady states}:
\begin{equation}
\delta\sigma (\dot\gamma) = b^{-x}f(\dot\gamma b^z)\enspace,
\end{equation}
where  $b$ is an arbitrary length rescaling factor, $f$ is the scaling function, and $x$ and $z$ are critical exponents. As $b$ is arbitrary, we may choose $b=\dot\gamma^{-1/z}$ in the above to get $\delta\sigma\sim \dot\gamma^{x/z}$, which is just the Herschel-Bulkley law with exponent $n=x/z$.  

To describe our present experiments, involving the {\it transient} decay to steady state, it is then natural to add time to the scaling equation above:
\begin{equation}
\delta\sigma (\dot\gamma, t)=b^{-x}f(\dot\gamma b^{z}, t b^{-z^{\prime}})\enspace.
\label{sl_sigma}
\end{equation}
Since the strain rate $\dot\gamma$ is an inverse time, it would be natural to assume $z=z^\prime$.  However, here we allow for the possibility of anomalous scaling for the ordering field $\dot\gamma$ and so we keep $z$ and $z^\prime$ distinct.  This will be necessary to describe our experimental results.  Again choosing $b=\dot\gamma^{-1/z}$ we get:
\begin{equation}
\delta\sigma (\dot\gamma, t)=\dot\gamma^{x/z}f(1, t \dot\gamma^{z^{\prime}/z})\enspace.
\label{sl_sigma2}
\end{equation}
For infinitely long times, the system reaches a steady state, the scaling function approaches a constant, and we recover the Herschel-Bulkley rheology.  The decay to this steady state is governed by the scaling variable $t \dot\gamma^{z^{\prime}/z}$, thus yielding the fluidization time under controlled shear rate, $\tau_f^{(\dot\gamma)}\sim \dot\gamma^{-\alpha}$, with $\alpha=z^{\prime}/z$. 

Symmetrically, one can impose a controlled stress and measure the time evolution of the shear rate. Its evolution is obtained by inverting Eq.~(\ref{sl_sigma}):
\begin{equation}
\dot\gamma(\delta\sigma , t)=b^{-z}g(\delta\sigma b^x, t b^{-z^{\prime}})\enspace,
\label{sl_gamma}
\end{equation}
where $g$ is a new scaling function.
In this case, it is convenient to choose $b=\delta\sigma^{-1/x}$ so that the evolution of the shear rate follows:
\begin{equation}
\dot\gamma(\delta\sigma , t)=\delta\sigma^{z/x}g(1, t\delta\sigma^{z^{\prime}/x})\enspace.
\label{sl_gamma2}
\end{equation}
As the long time steady state is approached, the scaling function $g$ approaches a constant, and one regains the same  expression for the Herschel-Bulkley law with exponent, $n=x/z$.  The approach to steady state is governed by the scaling variable $t\delta\sigma^{z^{\prime}/x}$, which determines the fluidization time under controlled stress,  $\tau_f^{(\sigma)}\sim \delta\sigma^{-\beta}$ with $\beta=z^{\prime}/x$. This scaling approach allows one to express the fluidization exponents $\alpha$ and $\beta$ in terms of the critical exponents $x$, $z$ and $z^\prime$. Moreover, one naturally finds that the ratio between the two fluidization exponents gives exactly the Herschel-Bulkley exponent:
\begin{equation}
\alpha/\beta=x/z=n\,,
\end{equation}
in perfect agreement with the experimental findings.\cite{Divoux:2011a}

The above connection with critical phenomena raises a number open issues that should be addressed in the future. First, from the theoretical point of view, is it possible to justify the scaling equations (\ref{sl_sigma}) and (\ref{sl_gamma}) based on rigorous grounds rather than on a simple analogy? One can indeed question whether these scaling equations apply to transient behavior so far from steady state, such as the fluidization process studied here. Second, the anomalous scaling with $z\neq z'$ should also be considered as well as the possibility for the exponents to take values that depend on the details of the jammed system. Experimentally, both the Herschel-Bulkley exponent $n$ and the fluidization exponents $\alpha$ and $\beta$ were shown to depend on various control parameters such as the carbopol concentration or the microgel pH. Since the volume fraction of soft jammed microgel particles is also a function of such control parameters, we suggest that the observed non-universality of the exponents might result from a dependence of critical properties on packing fraction (a ``line of critical points" above jamming) or perhaps from a qualitative change in dissipative processes that affects the critical behavior. Finally, the link between this very general approach in terms of critical phenomena and the analogy with delayed sedimentation discussed in Sect.~\ref{s.mechanism} to account for the induction-like phase also stands out as an open question. In other words, is there any connection between the ``critical'' state that the arrested microgel is assumed to reach right before the sudden fluidization at low shear rates and the critical-like scalings proposed above?

\section{Summary and conclusion}

In this paper, we have reported an extensive study of the transient shear-banding phenomenon observed in carbopol microgels during very long start-up experiments under controlled shear rate. By varying both the shear rate $\gp$ and the gap width $e$ in the concentric cylinder geometry, this data set completes our previous study \cite{Divoux:2010} and sheds new light on the fluidization dynamics of a Hershel-Bulkley fluid.

In particular, for low applied shear rates (typically smaller than 1~s$^{-1}$) or for small gap widths (typically smaller than 1~mm), the fluidization process involves the nucleation of shear band that remains almost stationary for several hours before large fluctuations give way to a homogeneous flow. This rather abrupt fluidization shows up on the stress response as a kink separating a quasi-stationary but fluctuating phase from full relaxation, which is also seen in the cone-and-plate geometry. For larger applied shear rates or for larger gap widths, the transient shear band is seen to continuously grow and invade the whole sample. In this case, no clear rheological signature can be associated with the fluidization process. The growth of the shear band is close to logarithmic in time, which is reminiscent of structural aging phenomena in soft glassy materials. We also conjectured that the state reached by the microgel before sudden fluidization is a kind of critical state similar to that observed in some delayed sedimentation experiments. 

Moreover, the fluidization time $\tau_f$, defined as the time after which the flow is homogeneously sheared, decreases as a power-law of the applied shear rate $\gp$, $\tau_f\sim \gp^{-\alpha}$ with an exponent $\alpha=1.8$ to 3.1, which mostly depends on the carbopol batch. In most cases, $\tau_f$ vs $\gp$ does not show any clear sign of the transition between the two above regimes for transient shear banding, although more experiments are still needed to draw definite conclusions on the ``phase diagram'' in the $(e,\gp)$ plane proposed in Sect.~\ref{s.phasediag}. We have also shown that boundary conditions seem to have a negligible influence on the long-time fluidization process. 
Finally, the general theoretical framework of critical phenomena was proposed to interpret the observed scaling laws for $\tau_f$. Such an approach nicely reconciles transient fluidization and steady-state Herschel-Bulkley rheology. It may also be relevant to the power-law behaviours reported in gels of type-I collagen \cite{Gobeaux:2010}.

The question remains whether the transient shear banding phenomenon explored here in carbopol microgels is shared by other simple yield stress fluids such as foams and emulsions. To provide an answer, future research should concentrate on revisiting previous studies with emphasis on small shear rates and on long start-up experiments. In particular, stress responses similar to those of Figs.~\ref{stress_seeding} and \ref{stress_geom} were reported on some emulsions together with transient heterogeneous flow.\cite{Becu:2005}. Transient shear banding regimes may also have been missed in previous studies due to the use of too large shear rates and/or too large gap widths. In other cases, it may have been wrongly interpreted in terms of stationary shear banding due to too short recording durations. The present study suggests that similar experiments should be conducted systematically on foams, emulsions and other microgels of better controlled microstructure \cite{Purnomo:2008}, following a rigorous rheological protocol so that the results obtained on different systems can be easily compared. Besides a possible universality in the fluidization processes of simple yield stress fluids, our results also raise the issue of the structure of the transient shear band: what are the microstructural differences, if any, between the fluidized and the solidlike material? To answer this question, experiments coupling time-resolved local visualization, e.g. through confocal microscopy, and rheometry are required on samples that allow for a fine tuning of the microstructural features, such as the size of jammed particles, their degree of softness, and their interactions.

\begin{acknowledgments}
We thank Y. Forterre for providing us with the carbopol, V.~Grenard for substantial help with the software, and H.~Feret for technical help. We also thank L.~Bocquet, A.~Colin, P.~Olsson, and G.~Ovarlez for several enlightening discussions as well as one anonymous referee for constructive comments.
\end{acknowledgments}

\bibliography{bibseb} 

\providecommand*{\mcitethebibliography}{\thebibliography}
\csname @ifundefined\endcsname{endmcitethebibliography}
{\let\endmcitethebibliography\endthebibliography}{}
\begin{mcitethebibliography}{69}
\providecommand*{\natexlab}[1]{#1}
\providecommand*{\mciteSetBstSublistMode}[1]{}
\providecommand*{\mciteSetBstMaxWidthForm}[2]{}
\providecommand*{\mciteBstWouldAddEndPuncttrue}
  {\def\EndOfBibitem{\unskip.}}
\providecommand*{\mciteBstWouldAddEndPunctfalse}
  {\let\EndOfBibitem\relax}
\providecommand*{\mciteSetBstMidEndSepPunct}[3]{}
\providecommand*{\mciteSetBstSublistLabelBeginEnd}[3]{}
\providecommand*{\EndOfBibitem}{}
\mciteSetBstSublistMode{f}
\mciteSetBstMaxWidthForm{subitem}
{(\emph{\alph{mcitesubitemcount}})}
\mciteSetBstSublistLabelBeginEnd{\mcitemaxwidthsubitemform\space}
{\relax}{\relax}

\bibitem[Barnes(1999)]{Barnes:1999}
H.~A. Barnes, \emph{J. Non-Newtonian Fluid Mech.}, 1999, \textbf{81},
  133--178\relax
\mciteBstWouldAddEndPuncttrue
\mciteSetBstMidEndSepPunct{\mcitedefaultmidpunct}
{\mcitedefaultendpunct}{\mcitedefaultseppunct}\relax
\EndOfBibitem
\bibitem[Weeks(2007)]{Weeks:2007}
E.~R. Weeks, Statistical Physics of Complex Fluids, 2007, pp. 243--255\relax
\mciteBstWouldAddEndPuncttrue
\mciteSetBstMidEndSepPunct{\mcitedefaultmidpunct}
{\mcitedefaultendpunct}{\mcitedefaultseppunct}\relax
\EndOfBibitem
\bibitem[H{\"o}hler and Addad(2005)]{Hohler:2005}
R.~H{\"o}hler and S.~C. Addad, \emph{J. Phys.: Condens. Matter}, 2005,
  \textbf{17}, 1041--1069\relax
\mciteBstWouldAddEndPuncttrue
\mciteSetBstMidEndSepPunct{\mcitedefaultmidpunct}
{\mcitedefaultendpunct}{\mcitedefaultseppunct}\relax
\EndOfBibitem
\bibitem[Schall. and van Hecke(2010)]{Schall:2010}
P.~Schall. and M.~van Hecke, \emph{Annu. Rev. Fluid Mech.}, 2010, \textbf{42},
  67--88\relax
\mciteBstWouldAddEndPuncttrue
\mciteSetBstMidEndSepPunct{\mcitedefaultmidpunct}
{\mcitedefaultendpunct}{\mcitedefaultseppunct}\relax
\EndOfBibitem
\bibitem[Coussot(2007)]{Coussot:2007}
P.~Coussot, \emph{Soft Matter}, 2007, \textbf{3}, 528--540\relax
\mciteBstWouldAddEndPuncttrue
\mciteSetBstMidEndSepPunct{\mcitedefaultmidpunct}
{\mcitedefaultendpunct}{\mcitedefaultseppunct}\relax
\EndOfBibitem
\bibitem[M{\o}ller \emph{et~al.}(2006)M{\o}ller, Mewis, and Bonn]{Moller:2006}
P.~C.~F. M{\o}ller, J.~Mewis and D.~Bonn, \emph{Soft Matter}, 2006, \textbf{2},
  274--283\relax
\mciteBstWouldAddEndPuncttrue
\mciteSetBstMidEndSepPunct{\mcitedefaultmidpunct}
{\mcitedefaultendpunct}{\mcitedefaultseppunct}\relax
\EndOfBibitem
\bibitem[Nguyen and Boger(1992)]{Nguyen:1992}
Q.~D. Nguyen and D.~V. Boger, \emph{Annu. Rev. Fluid Mech.}, 1992, \textbf{24},
  47--88\relax
\mciteBstWouldAddEndPuncttrue
\mciteSetBstMidEndSepPunct{\mcitedefaultmidpunct}
{\mcitedefaultendpunct}{\mcitedefaultseppunct}\relax
\EndOfBibitem
\bibitem[Roberts and Barnes(2001)]{Roberts:2001}
G.~P. Roberts and H.~A. Barnes, \emph{Rheol. Acta}, 2001, \textbf{40},
  499--503\relax
\mciteBstWouldAddEndPuncttrue
\mciteSetBstMidEndSepPunct{\mcitedefaultmidpunct}
{\mcitedefaultendpunct}{\mcitedefaultseppunct}\relax
\EndOfBibitem
\bibitem[M{\o}ller \emph{et~al.}(2009)M{\o}ller, Fall, and Bonn]{Moller:2009a}
P.~C.~F. M{\o}ller, A.~Fall and D.~Bonn, \emph{Europhys. Lett.}, 2009,
  \textbf{87}, 38004\relax
\mciteBstWouldAddEndPuncttrue
\mciteSetBstMidEndSepPunct{\mcitedefaultmidpunct}
{\mcitedefaultendpunct}{\mcitedefaultseppunct}\relax
\EndOfBibitem
\bibitem[Coussot \emph{et~al.}(2002)Coussot, Raynaud, Bertrand, Moucheront,
  Guilbaud, Huynh, Jarny, and Lesueur]{Coussot:2002a}
P.~Coussot, J.~S. Raynaud, F.~Bertrand, P.~Moucheront, J.~P. Guilbaud, H.~T.
  Huynh, S.~Jarny and D.~Lesueur, \emph{Phys. Rev. Lett.}, 2002, \textbf{88},
  218301\relax
\mciteBstWouldAddEndPuncttrue
\mciteSetBstMidEndSepPunct{\mcitedefaultmidpunct}
{\mcitedefaultendpunct}{\mcitedefaultseppunct}\relax
\EndOfBibitem
\bibitem[M{\o}ller \emph{et~al.}(2008)M{\o}ller, Rodts, Michels, and
  Bonn]{Moller:2008}
P.~C.~F. M{\o}ller, S.~Rodts, M.~A.~J. Michels and D.~Bonn, \emph{Phys. Rev.
  E}, 2008, \textbf{77}, 041507\relax
\mciteBstWouldAddEndPuncttrue
\mciteSetBstMidEndSepPunct{\mcitedefaultmidpunct}
{\mcitedefaultendpunct}{\mcitedefaultseppunct}\relax
\EndOfBibitem
\bibitem[Derec \emph{et~al.}(2001)Derec, Ajdari, and Lequeux]{Derec:2001}
C.~Derec, A.~Ajdari and F.~Lequeux, \emph{Eur. Phys. J. E}, 2001, \textbf{4},
  355--361\relax
\mciteBstWouldAddEndPuncttrue
\mciteSetBstMidEndSepPunct{\mcitedefaultmidpunct}
{\mcitedefaultendpunct}{\mcitedefaultseppunct}\relax
\EndOfBibitem
\bibitem[Picard \emph{et~al.}(2002)Picard, Ajdari, Bocquet, and
  Lequeux]{Picard:2002}
G.~Picard, A.~Ajdari, L.~Bocquet and F.~Lequeux, \emph{Phys. Rev. E}, 2002,
  \textbf{66}, 051501\relax
\mciteBstWouldAddEndPuncttrue
\mciteSetBstMidEndSepPunct{\mcitedefaultmidpunct}
{\mcitedefaultendpunct}{\mcitedefaultseppunct}\relax
\EndOfBibitem
\bibitem[Coussot and Ovarlez(2010)]{Coussot:2010}
P.~Coussot and G.~Ovarlez, \emph{Eur. Phys. J. E}, 2010, \textbf{33},
  183--188\relax
\mciteBstWouldAddEndPuncttrue
\mciteSetBstMidEndSepPunct{\mcitedefaultmidpunct}
{\mcitedefaultendpunct}{\mcitedefaultseppunct}\relax
\EndOfBibitem
\bibitem[Dennin(2008)]{Dennin:2008}
M.~Dennin, \emph{J. Phys.: Condens. Matter}, 2008, \textbf{20}, 283103\relax
\mciteBstWouldAddEndPuncttrue
\mciteSetBstMidEndSepPunct{\mcitedefaultmidpunct}
{\mcitedefaultendpunct}{\mcitedefaultseppunct}\relax
\EndOfBibitem
\bibitem[Mansard \emph{et~al.}(2011)Mansard, Collin, Chauduri, and
  Bocquet]{Mansard:2011}
V.~Mansard, A.~Collin, P.~Chauduri and L.~Bocquet, \emph{Soft Matter}, 2011,
  \textbf{7}, 5524--5527\relax
\mciteBstWouldAddEndPuncttrue
\mciteSetBstMidEndSepPunct{\mcitedefaultmidpunct}
{\mcitedefaultendpunct}{\mcitedefaultseppunct}\relax
\EndOfBibitem
\bibitem[Ragouilliaux \emph{et~al.}(2007)Ragouilliaux, Ovarlez,
  Shahidzadeh-Bonn, Herzhaft, Palermo, and Coussot]{Ragouilliaux:2007}
A.~Ragouilliaux, G.~Ovarlez, N.~Shahidzadeh-Bonn, B.~Herzhaft, T.~Palermo and
  P.~Coussot, \emph{Phys. Rev. E}, 2007, \textbf{76}, 051408\relax
\mciteBstWouldAddEndPuncttrue
\mciteSetBstMidEndSepPunct{\mcitedefaultmidpunct}
{\mcitedefaultendpunct}{\mcitedefaultseppunct}\relax
\EndOfBibitem
\bibitem[M{\o}ller \emph{et~al.}(2009)M{\o}ller, Fall, Chikkadi, Derks, and
  Bonn]{Moller:2009b}
P.~C.~F. M{\o}ller, A.~Fall, V.~Chikkadi, D.~Derks and D.~Bonn, \emph{Phil.
  Trans. R. Soc. Lond. A}, 2009, \textbf{367}, 5139--5155\relax
\mciteBstWouldAddEndPuncttrue
\mciteSetBstMidEndSepPunct{\mcitedefaultmidpunct}
{\mcitedefaultendpunct}{\mcitedefaultseppunct}\relax
\EndOfBibitem
\bibitem[Ovarlez \emph{et~al.}(2010)Ovarlez, Krishan, and Addad]{Ovarlez:2010}
G.~Ovarlez, K.~Krishan and S.~C. Addad, \emph{Europhys. Lett.}, 2010,
  \textbf{91}, 68005\relax
\mciteBstWouldAddEndPuncttrue
\mciteSetBstMidEndSepPunct{\mcitedefaultmidpunct}
{\mcitedefaultendpunct}{\mcitedefaultseppunct}\relax
\EndOfBibitem
\bibitem[Salmon \emph{et~al.}(2003)Salmon, B\'ecu, Manneville, and
  Colin]{Salmon:2003a}
J.-B. Salmon, L.~B\'ecu, S.~Manneville and A.~Colin, \emph{Eur. Phys. J. E},
  2003, \textbf{10}, 209--221\relax
\mciteBstWouldAddEndPuncttrue
\mciteSetBstMidEndSepPunct{\mcitedefaultmidpunct}
{\mcitedefaultendpunct}{\mcitedefaultseppunct}\relax
\EndOfBibitem
\bibitem[B{\'e}cu \emph{et~al.}(2006)B{\'e}cu, Manneville, and
  Colin]{Becu:2006}
L.~B{\'e}cu, S.~Manneville and A.~Colin, \emph{Phys. Rev. Lett.}, 2006,
  \textbf{96}, 138302\relax
\mciteBstWouldAddEndPuncttrue
\mciteSetBstMidEndSepPunct{\mcitedefaultmidpunct}
{\mcitedefaultendpunct}{\mcitedefaultseppunct}\relax
\EndOfBibitem
\bibitem[Rogers \emph{et~al.}(2008)Rogers, Vlassopoulos, and
  Callaghan]{Rogers:2008}
S.~A. Rogers, D.~Vlassopoulos and P.~T. Callaghan, \emph{Phys. Rev. Lett.},
  2008, \textbf{100}, 128304\relax
\mciteBstWouldAddEndPuncttrue
\mciteSetBstMidEndSepPunct{\mcitedefaultmidpunct}
{\mcitedefaultendpunct}{\mcitedefaultseppunct}\relax
\EndOfBibitem
\bibitem[Sollich \emph{et~al.}(1997)Sollich, Lequeux, H\'ebraud, and
  Cates]{Sollich:1997}
P.~Sollich, F.~Lequeux, P.~H\'ebraud and M.~E. Cates, \emph{Phys. Rev. Lett.},
  1997, \textbf{78}, 2020--2023\relax
\mciteBstWouldAddEndPuncttrue
\mciteSetBstMidEndSepPunct{\mcitedefaultmidpunct}
{\mcitedefaultendpunct}{\mcitedefaultseppunct}\relax
\EndOfBibitem
\bibitem[Sollich(1998)]{Sollich:1998}
P.~Sollich, \emph{Phys. Rev. E}, 1998, \textbf{58}, 738--759\relax
\mciteBstWouldAddEndPuncttrue
\mciteSetBstMidEndSepPunct{\mcitedefaultmidpunct}
{\mcitedefaultendpunct}{\mcitedefaultseppunct}\relax
\EndOfBibitem
\bibitem[Varnik \emph{et~al.}(2003)Varnik, Bocquet, Barrat, and
  Berthier]{Varnik:2003}
F.~Varnik, L.~Bocquet, J.-L. Barrat and L.~Berthier, \emph{Phys. Rev. Lett.},
  2003, \textbf{90}, 095702\relax
\mciteBstWouldAddEndPuncttrue
\mciteSetBstMidEndSepPunct{\mcitedefaultmidpunct}
{\mcitedefaultendpunct}{\mcitedefaultseppunct}\relax
\EndOfBibitem
\bibitem[Berthier(2003)]{Berthier:2003}
L.~Berthier, \emph{J. Phys.: Condens. Matter}, 2003, \textbf{15},
  S933--S943\relax
\mciteBstWouldAddEndPuncttrue
\mciteSetBstMidEndSepPunct{\mcitedefaultmidpunct}
{\mcitedefaultendpunct}{\mcitedefaultseppunct}\relax
\EndOfBibitem
\bibitem[Ovarlez \emph{et~al.}(2009)Ovarlez, Rodts, Chateau, and
  Coussot]{Ovarlez:2009}
G.~Ovarlez, S.~Rodts, X.~Chateau and P.~Coussot, \emph{Rheol. Acta}, 2009,
  \textbf{48}, 831--844\relax
\mciteBstWouldAddEndPuncttrue
\mciteSetBstMidEndSepPunct{\mcitedefaultmidpunct}
{\mcitedefaultendpunct}{\mcitedefaultseppunct}\relax
\EndOfBibitem
\bibitem[Mewis and Wagner(2009)]{Mewis:2009}
J.~Mewis and N.~Wagner, \emph{Adv. Colloid Interface Sci.}, 2009,
  \textbf{147--148}, 214--227\relax
\mciteBstWouldAddEndPuncttrue
\mciteSetBstMidEndSepPunct{\mcitedefaultmidpunct}
{\mcitedefaultendpunct}{\mcitedefaultseppunct}\relax
\EndOfBibitem
\bibitem[Besseling \emph{et~al.}(2010)Besseling, Isa, Ballesta, Petekidis,
  Cates, and Poon]{Besseling:2010}
R.~Besseling, L.~Isa, P.~Ballesta, G.~Petekidis, M.~E. Cates and W.~C.~K. Poon,
  \emph{Phys. Rev. Lett.}, 2010, \textbf{105}, 268301\relax
\mciteBstWouldAddEndPuncttrue
\mciteSetBstMidEndSepPunct{\mcitedefaultmidpunct}
{\mcitedefaultendpunct}{\mcitedefaultseppunct}\relax
\EndOfBibitem
\bibitem[Gibaud \emph{et~al.}(2008)Gibaud, Barentin, and
  Manneville]{Gibaud:2008}
T.~Gibaud, C.~Barentin and S.~Manneville, \emph{Phys. Rev. Lett.}, 2008,
  \textbf{101}, 258302\relax
\mciteBstWouldAddEndPuncttrue
\mciteSetBstMidEndSepPunct{\mcitedefaultmidpunct}
{\mcitedefaultendpunct}{\mcitedefaultseppunct}\relax
\EndOfBibitem
\bibitem[Gibaud \emph{et~al.}(2009)Gibaud, Barentin, Taberlet, and
  Manneville]{Gibaud:2009}
T.~Gibaud, C.~Barentin, N.~Taberlet and S.~Manneville, \emph{Soft Matter},
  2009, \textbf{5}, 3026--3037\relax
\mciteBstWouldAddEndPuncttrue
\mciteSetBstMidEndSepPunct{\mcitedefaultmidpunct}
{\mcitedefaultendpunct}{\mcitedefaultseppunct}\relax
\EndOfBibitem
\bibitem[Cheddadi \emph{et~al.}(2011)Cheddadi, Saramito, and
  Graner]{Cheddadi:2011}
I.~Cheddadi, P.~Saramito and F.~Graner, \emph{hal-006162273}, 2011\relax
\mciteBstWouldAddEndPuncttrue
\mciteSetBstMidEndSepPunct{\mcitedefaultmidpunct}
{\mcitedefaultendpunct}{\mcitedefaultseppunct}\relax
\EndOfBibitem
\bibitem[Divoux \emph{et~al.}(2010)Divoux, Tamarii, Barentin, and
  Manneville]{Divoux:2010}
T.~Divoux, D.~Tamarii, C.~Barentin and S.~Manneville, \emph{Phys. Rev. Lett.},
  2010, \textbf{104}, 208301\relax
\mciteBstWouldAddEndPuncttrue
\mciteSetBstMidEndSepPunct{\mcitedefaultmidpunct}
{\mcitedefaultendpunct}{\mcitedefaultseppunct}\relax
\EndOfBibitem
\bibitem[Divoux \emph{et~al.}(2011)Divoux, Barentin, and
  Manneville]{Divoux:2011b}
T.~Divoux, C.~Barentin and S.~Manneville, \emph{Soft Matter}, 2011, \textbf{7},
  9335--9349\relax
\mciteBstWouldAddEndPuncttrue
\mciteSetBstMidEndSepPunct{\mcitedefaultmidpunct}
{\mcitedefaultendpunct}{\mcitedefaultseppunct}\relax
\EndOfBibitem
\bibitem[Divoux \emph{et~al.}(2011)Divoux, Barentin, and
  Manneville]{Divoux:2011a}
T.~Divoux, C.~Barentin and S.~Manneville, \emph{Soft Matter}, 2011, \textbf{7},
  8409--8418\relax
\mciteBstWouldAddEndPuncttrue
\mciteSetBstMidEndSepPunct{\mcitedefaultmidpunct}
{\mcitedefaultendpunct}{\mcitedefaultseppunct}\relax
\EndOfBibitem
\bibitem[Katgert \emph{et~al.}(2009)Katgert, A.~Latka, and van
  Hecke]{Katgert:2009}
G.~Katgert, M.~E.~M. A.~Latka and M.~van Hecke, \emph{Phys. Rev. E}, 2009,
  \textbf{79}, 066318\relax
\mciteBstWouldAddEndPuncttrue
\mciteSetBstMidEndSepPunct{\mcitedefaultmidpunct}
{\mcitedefaultendpunct}{\mcitedefaultseppunct}\relax
\EndOfBibitem
\bibitem[Gobeaux \emph{et~al.}(2010)Gobeaux, Belamie, Mosser, Davidson, and
  Asnacios]{Gobeaux:2010}
F.~Gobeaux, E.~Belamie, G.~Mosser, P.~Davidson and S.~Asnacios, \emph{Soft
  Matter}, 2010, \textbf{6}, 3769--3777\relax
\mciteBstWouldAddEndPuncttrue
\mciteSetBstMidEndSepPunct{\mcitedefaultmidpunct}
{\mcitedefaultendpunct}{\mcitedefaultseppunct}\relax
\EndOfBibitem
\bibitem[Baudonnet \emph{et~al.}(2004)Baudonnet, Grossiord, and
  Rodriguez]{Baudonnet:2004}
L.~Baudonnet, J.-L. Grossiord and F.~Rodriguez, \emph{J. Dispersion Sci.
  Technol.}, 2004, \textbf{25}, 183--192\relax
\mciteBstWouldAddEndPuncttrue
\mciteSetBstMidEndSepPunct{\mcitedefaultmidpunct}
{\mcitedefaultendpunct}{\mcitedefaultseppunct}\relax
\EndOfBibitem
\bibitem[Ketz \emph{et~al.}(1988)Ketz, Prud'homme, and Graessley]{Ketz:1988}
R.~J. Ketz, R.~K. Prud'homme and W.~W. Graessley, \emph{Rheol. Acta}, 1988,
  \textbf{27}, 531--539\relax
\mciteBstWouldAddEndPuncttrue
\mciteSetBstMidEndSepPunct{\mcitedefaultmidpunct}
{\mcitedefaultendpunct}{\mcitedefaultseppunct}\relax
\EndOfBibitem
\bibitem[Kim \emph{et~al.}(2003)Kim, Song, Lee, and Park]{Kim:2003}
J.-Y. Kim, J.-Y. Song, E.-J. Lee and S.-K. Park, \emph{Colloid Polym. Sci.},
  2003, \textbf{281}, 614--623\relax
\mciteBstWouldAddEndPuncttrue
\mciteSetBstMidEndSepPunct{\mcitedefaultmidpunct}
{\mcitedefaultendpunct}{\mcitedefaultseppunct}\relax
\EndOfBibitem
\bibitem[Oppong \emph{et~al.}(2006)Oppong, Rubatat, Bailey, Frisken, and
  de~Bruyn]{Oppong:2006}
F.~K. Oppong, L.~Rubatat, A.~E. Bailey, B.~J. Frisken and J.~R. de~Bruyn,
  \emph{Phys. Rev. E}, 2006, \textbf{73}, 041405\relax
\mciteBstWouldAddEndPuncttrue
\mciteSetBstMidEndSepPunct{\mcitedefaultmidpunct}
{\mcitedefaultendpunct}{\mcitedefaultseppunct}\relax
\EndOfBibitem
\bibitem[Lee \emph{et~al.}(2011)Lee, Gutowski, Bailey, Rubatat, de~Bruyn, and
  Frisken]{Lee:2011}
D.~Lee, I.~A. Gutowski, A.~E. Bailey, L.~Rubatat, J.~R. de~Bruyn and B.~J.
  Frisken, \emph{Phys. Rev. E}, 2011, \textbf{83}, 031401\relax
\mciteBstWouldAddEndPuncttrue
\mciteSetBstMidEndSepPunct{\mcitedefaultmidpunct}
{\mcitedefaultendpunct}{\mcitedefaultseppunct}\relax
\EndOfBibitem
\bibitem[Piau(2007)]{Piau:2007}
J.~M. Piau, \emph{J. Non-Newtonian Fluid Mech.}, 2007, \textbf{144},
  1--29\relax
\mciteBstWouldAddEndPuncttrue
\mciteSetBstMidEndSepPunct{\mcitedefaultmidpunct}
{\mcitedefaultendpunct}{\mcitedefaultseppunct}\relax
\EndOfBibitem
\bibitem[Coussot \emph{et~al.}(2009)Coussot, Tocquer, Lanos, and
  Ovarlez]{Coussot:2009}
P.~Coussot, L.~Tocquer, C.~Lanos and G.~Ovarlez, \emph{J. Non-Newtonian Fluid
  Mech.}, 2009, \textbf{158}, 85--90\relax
\mciteBstWouldAddEndPuncttrue
\mciteSetBstMidEndSepPunct{\mcitedefaultmidpunct}
{\mcitedefaultendpunct}{\mcitedefaultseppunct}\relax
\EndOfBibitem
\bibitem[Benmouffok-Benbelkacem \emph{et~al.}(2010)Benmouffok-Benbelkacem,
  Caton, Baravian, and Skali-Lami]{Benmouffok:2010}
G.~Benmouffok-Benbelkacem, F.~Caton, C.~Baravian and S.~Skali-Lami,
  \emph{Rheol. Acta}, 2010, \textbf{49}, 305--314\relax
\mciteBstWouldAddEndPuncttrue
\mciteSetBstMidEndSepPunct{\mcitedefaultmidpunct}
{\mcitedefaultendpunct}{\mcitedefaultseppunct}\relax
\EndOfBibitem
\bibitem[Manneville \emph{et~al.}(2004)Manneville, B{\'e}cu, and
  Colin]{Manneville:2004a}
S.~Manneville, L.~B{\'e}cu and A.~Colin, \emph{Eur. Phys. J. AP}, 2004,
  \textbf{28}, 361--373\relax
\mciteBstWouldAddEndPuncttrue
\mciteSetBstMidEndSepPunct{\mcitedefaultmidpunct}
{\mcitedefaultendpunct}{\mcitedefaultseppunct}\relax
\EndOfBibitem
\bibitem[Cloitre \emph{et~al.}(2000)Cloitre, Borrega, and
  Leibler]{Cloitre:2000}
M.~Cloitre, R.~Borrega and L.~Leibler, \emph{Phys. Rev. Lett.}, 2000,
  \textbf{85}, 4819--4822\relax
\mciteBstWouldAddEndPuncttrue
\mciteSetBstMidEndSepPunct{\mcitedefaultmidpunct}
{\mcitedefaultendpunct}{\mcitedefaultseppunct}\relax
\EndOfBibitem
\bibitem[Viasnoff and Lequeux(2002)]{Viasnoff:2002}
V.~Viasnoff and F.~Lequeux, \emph{Phys. Rev. Lett.}, 2002, \textbf{89},
  065701\relax
\mciteBstWouldAddEndPuncttrue
\mciteSetBstMidEndSepPunct{\mcitedefaultmidpunct}
{\mcitedefaultendpunct}{\mcitedefaultseppunct}\relax
\EndOfBibitem
\bibitem[Goyon \emph{et~al.}(2008)Goyon, Colin, Ovarlez, Ajdari, and
  Bocquet]{Goyon:2008}
J.~Goyon, A.~Colin, G.~Ovarlez, A.~Ajdari and L.~Bocquet, \emph{Nature}, 2008,
  \textbf{454}, 84--87\relax
\mciteBstWouldAddEndPuncttrue
\mciteSetBstMidEndSepPunct{\mcitedefaultmidpunct}
{\mcitedefaultendpunct}{\mcitedefaultseppunct}\relax
\EndOfBibitem
\bibitem[Ovarlez \emph{et~al.}(2008)Ovarlez, Rodts, Ragouilliaux, Coussot,
  Goyon, and Colin]{Ovarlez:2008}
G.~Ovarlez, S.~Rodts, A.~Ragouilliaux, P.~Coussot, J.~Goyon and A.~Colin,
  \emph{Phys. Rev. E}, 2008, \textbf{78}, 036307\relax
\mciteBstWouldAddEndPuncttrue
\mciteSetBstMidEndSepPunct{\mcitedefaultmidpunct}
{\mcitedefaultendpunct}{\mcitedefaultseppunct}\relax
\EndOfBibitem
\bibitem[Huang \emph{et~al.}(2005)Huang, Ovarlez, Bertrand, Rodts, Coussot, and
  Bonn]{Huang:2005}
N.~Huang, G.~Ovarlez, F.~Bertrand, S.~Rodts, P.~Coussot and D.~Bonn,
  \emph{Phys. Rev. Lett.}, 2005, \textbf{94}, 028301\relax
\mciteBstWouldAddEndPuncttrue
\mciteSetBstMidEndSepPunct{\mcitedefaultmidpunct}
{\mcitedefaultendpunct}{\mcitedefaultseppunct}\relax
\EndOfBibitem
\bibitem[Ragouilliaux \emph{et~al.}(2006)Ragouilliaux, Herzhaft, Bertrand, and
  Coussot]{Ragouilliaux:2006}
A.~Ragouilliaux, B.~Herzhaft, F.~Bertrand and P.~Coussot, \emph{Rheol. Acta},
  2006, \textbf{46}, 261--271\relax
\mciteBstWouldAddEndPuncttrue
\mciteSetBstMidEndSepPunct{\mcitedefaultmidpunct}
{\mcitedefaultendpunct}{\mcitedefaultseppunct}\relax
\EndOfBibitem
\bibitem[Meeker \emph{et~al.}(2004)Meeker, Bonnecaze, and
  Cloitre]{Meeker:2004a}
S.~P. Meeker, R.~T. Bonnecaze and M.~Cloitre, \emph{Phys. Rev. Lett.}, 2004,
  \textbf{92}, 198302\relax
\mciteBstWouldAddEndPuncttrue
\mciteSetBstMidEndSepPunct{\mcitedefaultmidpunct}
{\mcitedefaultendpunct}{\mcitedefaultseppunct}\relax
\EndOfBibitem
\bibitem[Meeker \emph{et~al.}(2004)Meeker, Bonnecaze, and
  Cloitre]{Meeker:2004b}
S.~P. Meeker, R.~T. Bonnecaze and M.~Cloitre, \emph{J. Rheol.}, 2004,
  \textbf{48}, 1295--1320\relax
\mciteBstWouldAddEndPuncttrue
\mciteSetBstMidEndSepPunct{\mcitedefaultmidpunct}
{\mcitedefaultendpunct}{\mcitedefaultseppunct}\relax
\EndOfBibitem
\bibitem[Goyon \emph{et~al.}(2010)Goyon, Colin, and Bocquet]{Goyon:2010}
J.~Goyon, A.~Colin and L.~Bocquet, \emph{Soft Matter}, 2010, \textbf{6},
  2668--2678\relax
\mciteBstWouldAddEndPuncttrue
\mciteSetBstMidEndSepPunct{\mcitedefaultmidpunct}
{\mcitedefaultendpunct}{\mcitedefaultseppunct}\relax
\EndOfBibitem
\bibitem[Bocquet \emph{et~al.}(1998)Bocquet, Charlaix, Ciliberto, and
  Crassous]{Bocquet:1998}
L.~Bocquet, E.~Charlaix, S.~Ciliberto and J.~Crassous, \emph{Nature}, 1998,
  \textbf{396}, 735--737\relax
\mciteBstWouldAddEndPuncttrue
\mciteSetBstMidEndSepPunct{\mcitedefaultmidpunct}
{\mcitedefaultendpunct}{\mcitedefaultseppunct}\relax
\EndOfBibitem
\bibitem[Gayvallet and G{\'e}minard(2002)]{Gayvallet:2002}
H.~Gayvallet and J.-C. G{\'e}minard, \emph{Eur. Phys. J. B}, 2002, \textbf{30},
  369--375\relax
\mciteBstWouldAddEndPuncttrue
\mciteSetBstMidEndSepPunct{\mcitedefaultmidpunct}
{\mcitedefaultendpunct}{\mcitedefaultseppunct}\relax
\EndOfBibitem
\bibitem[Ronsin \emph{et~al.}(2009)Ronsin, Carolin, and
  Baumberger]{Ronsin:2009}
O.~Ronsin, C.~Carolin and T.~Baumberger, \emph{Phys. Rev. Lett.}, 2009,
  \textbf{103}, 138302\relax
\mciteBstWouldAddEndPuncttrue
\mciteSetBstMidEndSepPunct{\mcitedefaultmidpunct}
{\mcitedefaultendpunct}{\mcitedefaultseppunct}\relax
\EndOfBibitem
\bibitem[Barlett \emph{et~al.}(2011)Barlett, Teece, and Faers]{Barlett:2011}
P.~Barlett, L.~Teece and M.~Faers, \emph{ArXiv:1109.4893v1 [cond-mat.soft].},
  2011\relax
\mciteBstWouldAddEndPuncttrue
\mciteSetBstMidEndSepPunct{\mcitedefaultmidpunct}
{\mcitedefaultendpunct}{\mcitedefaultseppunct}\relax
\EndOfBibitem
\bibitem[Poon \emph{et~al.}(1999)Poon, Starrs, Meeker, Moussad, Evans, Pusey,
  and Robins]{Poon:1999}
W.~C.~K. Poon, L.~Starrs, S.~P. Meeker, A.~Moussad, R.~M.~L. Evans, P.~N. Pusey
  and M.~M. Robins, \emph{Faraday Discuss.}, 1999, \textbf{112}, 143--154\relax
\mciteBstWouldAddEndPuncttrue
\mciteSetBstMidEndSepPunct{\mcitedefaultmidpunct}
{\mcitedefaultendpunct}{\mcitedefaultseppunct}\relax
\EndOfBibitem
\bibitem[Manley \emph{et~al.}(2005)Manley, Skotheim, Mahadevan, and
  Weitz]{Manley:2005}
S.~Manley, J.~M. Skotheim, L.~Mahadevan and D.~A. Weitz, \emph{Phys. Rev.
  Lett.}, 2005, \textbf{94}, 218302\relax
\mciteBstWouldAddEndPuncttrue
\mciteSetBstMidEndSepPunct{\mcitedefaultmidpunct}
{\mcitedefaultendpunct}{\mcitedefaultseppunct}\relax
\EndOfBibitem
\bibitem[Testard \emph{et~al.}(2011)Testard, Berthier, and Kob]{Testard:2011}
V.~Testard, L.~Berthier and W.~Kob, \emph{Phys. Rev. Lett.}, 2011,
  \textbf{106}, 125702\relax
\mciteBstWouldAddEndPuncttrue
\mciteSetBstMidEndSepPunct{\mcitedefaultmidpunct}
{\mcitedefaultendpunct}{\mcitedefaultseppunct}\relax
\EndOfBibitem
\bibitem[Purnomo \emph{et~al.}(2008)Purnomo, van~den Ende, Vanapalli, and
  Mugele]{Purnomo:2008}
E.~H. Purnomo, D.~van~den Ende, S.~A. Vanapalli and F.~Mugele, \emph{Phys. Rev.
  Lett.}, 2008, \textbf{101}, 238301\relax
\mciteBstWouldAddEndPuncttrue
\mciteSetBstMidEndSepPunct{\mcitedefaultmidpunct}
{\mcitedefaultendpunct}{\mcitedefaultseppunct}\relax
\EndOfBibitem
\bibitem[Coussot \emph{et~al.}(2006)Coussot, Tabuteau, Chateau, Tocquer, and
  Ovarlez]{Coussot:2006}
P.~Coussot, H.~Tabuteau, X.~Chateau, L.~Tocquer and G.~Ovarlez, \emph{J.
  Rheol.}, 2006, \textbf{50}, 975--994\relax
\mciteBstWouldAddEndPuncttrue
\mciteSetBstMidEndSepPunct{\mcitedefaultmidpunct}
{\mcitedefaultendpunct}{\mcitedefaultseppunct}\relax
\EndOfBibitem
\bibitem[Joshi and Reddy(2008)]{Joshi:2008b}
Y.~M. Joshi and G.~R.~K. Reddy, \emph{Phys. Rev. E}, 2008, \textbf{77},
  021501\relax
\mciteBstWouldAddEndPuncttrue
\mciteSetBstMidEndSepPunct{\mcitedefaultmidpunct}
{\mcitedefaultendpunct}{\mcitedefaultseppunct}\relax
\EndOfBibitem
\bibitem[Lema{\^i}tre and Caroli(2009)]{Lemaitre:2009}
A.~Lema{\^i}tre and C.~Caroli, \emph{Phys. Rev. Lett.}, 2009, \textbf{103},
  065501\relax
\mciteBstWouldAddEndPuncttrue
\mciteSetBstMidEndSepPunct{\mcitedefaultmidpunct}
{\mcitedefaultendpunct}{\mcitedefaultseppunct}\relax
\EndOfBibitem
\bibitem[Martens \emph{et~al.}(2011)Martens, Bocquet, and Barrat]{Martens:2011}
K.~Martens, L.~Bocquet and J.-L. Barrat, \emph{Phys. Rev. Lett.}, 2011,
  \textbf{106}, 165001\relax
\mciteBstWouldAddEndPuncttrue
\mciteSetBstMidEndSepPunct{\mcitedefaultmidpunct}
{\mcitedefaultendpunct}{\mcitedefaultseppunct}\relax
\EndOfBibitem
\bibitem[Wiley and Sons(1975)]{Barton:1975}
\emph{Introduction to renormalization group and critical phenomena}, ed.
  J.~Wiley and Sons, 1975\relax
\mciteBstWouldAddEndPuncttrue
\mciteSetBstMidEndSepPunct{\mcitedefaultmidpunct}
{\mcitedefaultendpunct}{\mcitedefaultseppunct}\relax
\EndOfBibitem
\bibitem[B{\'e}cu \emph{et~al.}(2005)B{\'e}cu, Grondin, Manneville, and
  Colin]{Becu:2005}
L.~B{\'e}cu, P.~Grondin, S.~Manneville and A.~Colin, \emph{Colloids Surfaces
  A}, 2005, \textbf{263}, 146--152\relax
\mciteBstWouldAddEndPuncttrue
\mciteSetBstMidEndSepPunct{\mcitedefaultmidpunct}
{\mcitedefaultendpunct}{\mcitedefaultseppunct}\relax
\EndOfBibitem
\end{mcitethebibliography}
\bibliographystyle{rsc} 

\clearpage

\vspace{0.6cm}

\section{Supplementary Figure 1}

The total slip velocity $v_s$ was computed as the sum of slip velocities at the rotor and at the stator recorded once steady state is reached and averaged over at least 150~s. Supplementary Figure~1 shows  $v_s$ relative to the rotor velocity $v_0$ for various gap widths and applied shear rates. $v_s/v_0$ is much larger for small gap widths than for larger gaps. For $e=0.45$~mm, the amount of wall slip is about 30~\%, independent of or weakly decreasing with $\gp$. The same trend is observed for $e=1.5$~mm yet wall slip is smaller and of the order of 15~\%. For $e=3$~mm however, the relative slip sharply decreases with $\gp$ from about 20~\% at the lowest shear rate down to negligible values of the order of our uncertainty of about 2~\%. 

\begin{figure}[!h]\tt
\centering
\includegraphics[width=0.9\columnwidth]{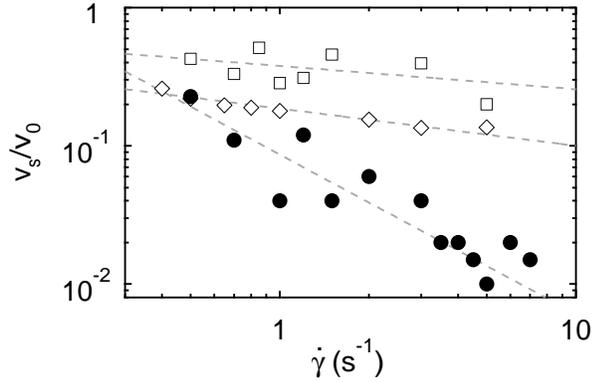}
\caption{Steady-state total slip velocity $v_s$ relative to the rotor velocity $v_0$ as a function of the applied shear rate $\gp$ for smooth Couette geometries of gap width $e=0.45$~mm ($\square$), $e=1.5$~mm ($\lozenge$), and $e=3$~mm ($\bullet$). Dotted lines are power laws drawn to guide the eye. Experiments performed on different batches of 1~\%~w/w carbopol microgels.}
\label{slipvel}
\end{figure}

\section{Supplementary Figure 2}

\begin{figure*}[!h]\tt
\centering
\includegraphics[width=0.8\linewidth]{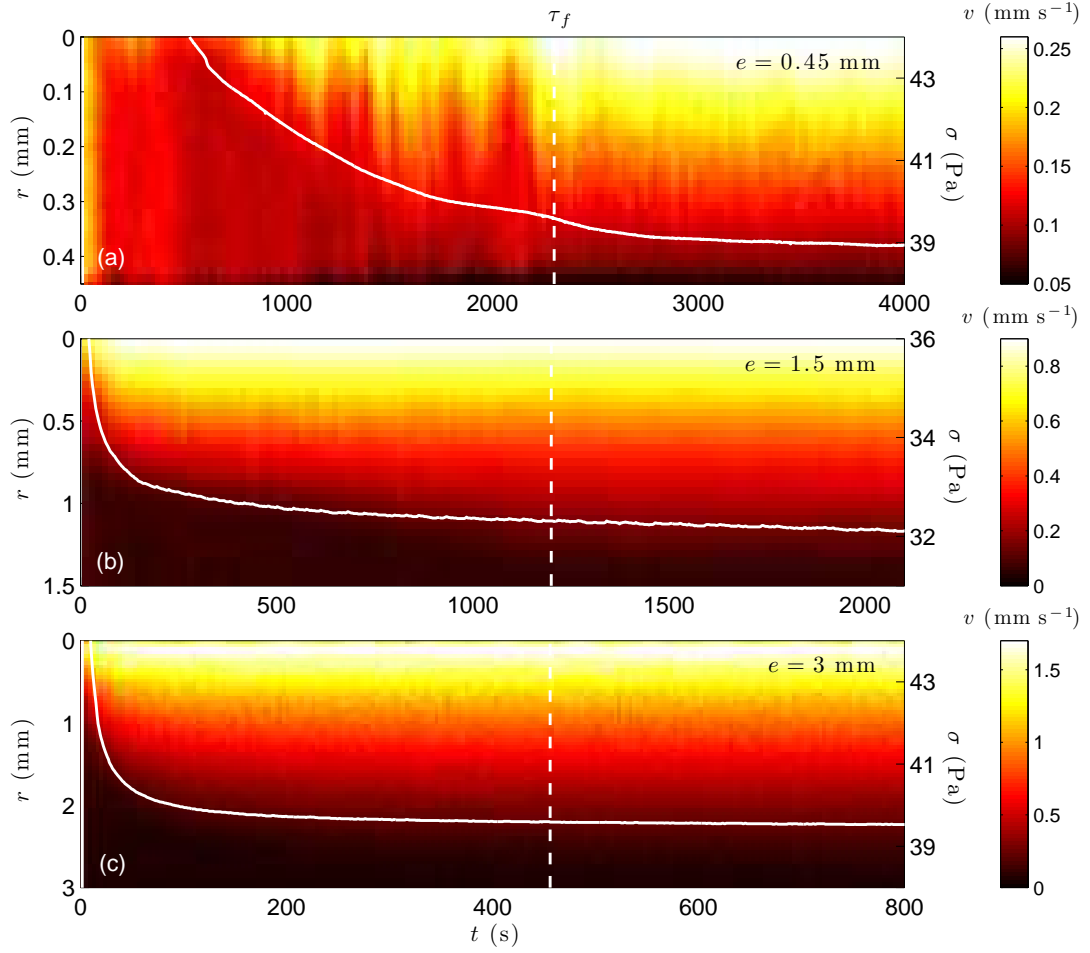}
\caption{Spatiotemporal diagrams of the velocity data $v(r,t)$ in smooth Couette cells of different gap widths $e$ and under similar shear rates.  (a)~$e=0.45$~mm and $\gp=0.85$~s$^{-1}$. (b)~$e=1.5$~mm and $\gp=0.8$~s$^{-1}$. (c)~$e=3$~mm and $\gp=0.7$~s$^{-1}$. White lines are the corresponding stress responses $\sigma(t)$ (right vertical axis). The vertical dashed lines indicate the fluidization times $\tau_f$. The time interval between two velocity profiles is 21~s, 17~s, and 4~s in (a), (b), and (c) respectively. }
\label{sptp_gaps}
\end{figure*}

Supplementary Figure~2 shows the spatiotemporal diagrams of $v(x,t)$ recorded in the three different Couette cells for similar shear rates $\gp\simeq 0.8$~s$^{-1}$ together with the corresponding stress responses. Transient shear banding is observed for $t<\tau_f$ as indicated by white dotted lines (see also the movies in the supplementary material\dag). 

The results for $e=0.45$~mm are qualitatively similar to those found previously for $e=1$~mm and $\gp<\gps$ [see e.g. $\gp=1.5$~s$^{-1}$ in Fig.~9(a)], i.e. the transient regime presents a quasi-stationary phase followed by strong fluctuations and abrupt full fluidization. Here, it is interesting to note that the quasi-stationary phase ($t\lesssim 10^3$~s) involves a pluglike flow at about half the rotor velocity whose velocity slowly decreases and that precedes the nucleation of a fluctuating shear band (see also the movie in the supplementary material\dag). This initial plug flow is only seen in smooth geometries and was already evidenced in a previous study devoted to the stress overshoot phenomenon at short times.\cite{Divoux:2011b}

On the other hand, the spatiotemporal diagrams for $e=1.5$~mm and $e=3$~mm resemble that of Fig.~9(c). Indeed, as the gap width is increased, the characteristic kink in $\sigma(t)$ disappears as well as the fluctuations of the flow field, and the shear banding regime becomes more progressive and continuous. 

\end{document}